\documentstyle[psfig,aps,prl]{revtex}
\begin{document}
\title{Quantum conditional operator and a criterion for separability}
\author{N. J. Cerf$^{1,2}$, C. Adami$^1$, and R. M. Gingrich$^1$}
\address{$^1$W. K. Kellogg Radiation Laboratory, 
California Institute of Technology, Pasadena, California 91125\\
$^2$Information Systems Technology Section, Jet Propulsion Laboratory,
Pasadena, California 91109}

\date{September 1997}

\draft
\maketitle
\vskip -0.1cm

\begin{abstract}
We analyze the properties of the conditional {\em amplitude} operator, 
the quantum analog of the conditional probability which has been
introduced in [quant-ph/9512022]. 
The spectrum of the conditional operator characterizing a
quantum bipartite system is invariant under local unitary
transformations and reflects its inseparability. More
specifically, it is shown that the conditional amplitude operator
of a separable state cannot have an eigenvalue exceeding 1, 
which results in a necessary condition for separability.
This leads us to consider a related separability criterion
based on the positive map $\Gamma:\rho \to ({\rm Tr} \rho) - \rho$, where
$\rho$ is an Hermitian operator. Any separable state is mapped by the
tensor product of this map and the identity into a non-negative
operator, which provides a simple necessary condition for
separability. In the special case where one subsystem is a
quantum bit, $\Gamma$ reduces to time-reversal, 
so that this separability condition is equivalent to partial transposition.
It is therefore also sufficient for $2\times 2$ and $2\times 3$ systems.
Finally, a simple connection between this map and complex conjugation 
in the ``magic'' basis is displayed. 
\end{abstract}

\pacs{PACS numbers: 03.65.Bz, 89.70.+c
      \hfill KRL preprint MAP-217}

\section{Introduction}

The state of a quantum bipartite system $AB$ is 
described as {\em separable} (or classically correlated)
if it can be obtained by having both
parties $A$ and $B$ preparing their subsystem according to some
common instructions (see, e.g., \cite{bib_werner,bib_peres}). 
Mathematically, this means that the density
operator $\rho$ characterizing the state of the bipartite
system can be written as a convex sum of product states, that is
\begin{equation}  \label{eq_simple}
\rho = \sum_i w_i \left( \rho_i^{(A)} \otimes \rho_i^{(B)} \right)
\end{equation}
where the weights $w_i$ satisfy $\sum_i w_i=1$ and $0\le w_i \le 1$.
The $w_i$'s can be viewed as the probability distribution of 
a classical random variable that is known to both parties $A$ and $B$
and used by them to prepare their subsystem. Namely,
if the subsystem $A$ (and $B$) is prepared in state $\rho_i^{(A)}$
(and $\rho_i^{(B)}$) when the classical variable takes on value $i$,
the state of the joint system $AB$ is given by Eq.~(\ref{eq_simple}).
A separable state $\rho$ satisfies several interesting properties.
The joint statistics of any pair of {\em local}
observables $O_A$ and $O_B$ (measured separately on each subsystem)
can be described classically, based on an underlying global
``hidden'' variable. For example, the quantum expectation value 
of the product $O_A O_B$ is given by
\begin{equation}
{\rm Tr}[\rho (O_A \otimes O_B)] = \sum_i w_i 
\langle a\rangle_i \langle b\rangle_i
\end{equation}
where $\langle a\rangle_i={\rm Tr}[\rho_i^{(A)} O_A]$ and 
$\langle b\rangle_i={\rm Tr}[\rho_i^{(B)} O_B]$. In other words, 
the joint statistics of $O_A$ and $O_B$ can be understood classically,
by assuming that the local statistics of the outcomes
can be described separately for each
$\rho_i^{(A)}$ and  $\rho_i^{(B)}$, and that the correlations
originate from a hidden variable $i$ distributed
according to $w_i$. Moreover, a separable system always satisfies Bell's
inequalities (the converse is not true), 
so that the latter represent a {\em necessary}
condition for separability (see, e.g., \cite{bib_werner}). 
Note that any joint probability distribution can be represented
as a convex combination of product distributions, so that classical
probabilities are always separable in the above sense.
\par

The decomposition of a separable state $\rho$ 
into a convex mixture of product states 
is not unique in general, but the fact that $\rho$ is separable
implies that there must exist at least one such decomposition.
If no such decomposition can be written, then $\rho$ is termed
{\em inseparable}, and it can be viewed as {\em quantum} correlated. 
Except for the special case where $\rho$ describes a pure state,
the distinction between separable and inseparable states 
appears to be an extraordinary difficult problem. 
More precisely, some mixed states can be ``weakly'' inseparable, 
in the sense that it is very hard
to establish with certainty their inseparability. This is basically
due to the difficulty of enumerating explicitly {\em all} the possible
convex combinations of product states in order to detect 
that a state is actually inseparable. 
Still, it is possible to find some
conditions that {\em all} separable states must satisfy, therefore
allowing the detection of inseparability when a state violates
one such condition. The most common example of such a {\em necessary} 
condition for separability is the satisfaction of Bell's inequalities. A
state that violates Bell's inequalities is inseparable, while a state
satisfying them may be separable or weakly inseparable~\cite{bib_werner}.
\par

More recently, a surprisingly simple {\em necessary} condition 
for separability has been discovered by Peres~\cite{bib_peres},
which has been shown by Horodecki et al.~\cite{bib_horo_sufficient}
to be strong enough to {\em guarantee} separability for bipartite
systems of dimension $2\times 2$ and $2\times3$.
If the state
$\rho$ is separable, then the operator obtained by applying 
a {\em partial} transposition with respect to subsystem $A$ (or $B$) to
$\rho$ must be positive, that is
\begin{equation}
\rho^{T_A} = \left( \rho^{T_B} \right)^* \ge 0
\end{equation}
Thus, this criterion amounts to checking that all the eigenvalues
of the partial transposition of $\rho$ are non-negative, which
must be so for all separable states.
In Hilbert spaces of dimensions $2\times 2$ and $2\times 3$, 
this condition is actually {\em sufficient}, that is, it suffices for
ruling out {\em all} inseparable states~\cite{bib_horo_sufficient}. 
In larger dimensions, however, it
is provably {\em not} sufficient, in the sense that it does not
detect some weakly inseparable 
states~\cite{bib_horo_sufficient,bib_horo_one}.
A general necessary {\em and} sufficient condition
for separability in arbitrary dimensions has been found
by Horodecki et al.~\cite{bib_horo_sufficient}, 
which states that $\rho$ is separable if and only
if the tensor product of {\em any} positive\footnote{A map is defined
as positive if it maps positive operators into positive operators.} 
map (acting on $A$) and the identity (acting on $B$)
maps $\rho$ into a positive operator. Although very important
in theory, this criterion is hardly more practical than the definition
of separability itself since it involves the characterization of the
set of all positive maps. It appears to be useful mainly
for $2\times 2$ and $2\times 3$ bipartite systems, where such a general 
characterization has been found~\cite{bib_horo_sufficient}.
\par

In this paper, we focus on the connection between quantum non-separability
and the conditional amplitude operator which has been 
introduced recently in the context of 
quantum information theory~\cite{bib_neginfo,bib_oviedo,bib_physcomp}. 
In Section II, we start by detailing the mathematical properties
of such an operator (support, spectrum, connection
with von Neumann entropies, etc.).
We then derive a {\em necessary} condition for separability, 
based on the conditional von Neumann entropy and 
its underlying conditional amplitude operator. Namely, the eigenvalues
of the latter operator cannot exceed 1 if the bipartite state
is separable, as was conjectured in 
Refs.~\cite{bib_neginfo,bib_oviedo,bib_physcomp}.
Since the conditional von Neumann entropy can be negative only if
the conditional amplitude operator admits an eigenvalue larger than 1,
a related---and weaker---separability condition is
that the conditional entropy is non-negative 
(see also Refs.~\cite{bib_neginfo,bib_oviedo,bib_physcomp}
and \cite{bib_hor_alphaentropy}). 
This leads us to consider a positive map 
$\Gamma: \rho \to ({\rm Tr}\rho)-\rho$
which gives rise to a simple {\em necessary} condition for separability
in arbitrary dimensions. More specifically, it is shown 
in Section III that any separable
state is mapped by the tensor product of $\Gamma$ (acting on one
subsystem, $A$) and the identity (acting on the other, $B$)
into a non-negative operator.
In the case where $\Gamma$ is applied to a two-state system 
(quantum bit or spin-1/2
particle), this corresponds to the time-reversal operation
applied on one system with respect to the other one. Since
Peres' criterion has been shown to be unitarily equivalent 
to such a ``local'' time-reversal by Sanpera et al.~\cite{bib_sanpera}, 
our separability criterion is simply
equivalent to Peres' for $2\times n$ composite systems. 
Therefore, it also results in a {\em sufficient} condition
for $2\times 2$ and $2\times 3$ systems, according to
Ref.~\cite{bib_horo_sufficient}.
It also has a very simple geometric
representation in the Hilbert-Schmidt representation of the bipartite
state. Finally, it appears that the map $\Gamma$ is connected
to the complex conjugation operation in the ``magic'' basis
introduced by Hill and Wootters~\cite{bib_hillwootters}.
In Appendix A, we illustrate the separability condition based on $\Gamma$
by applying it to several separable or inseparable states, and compare
it to the separability condition based on partial transposition.
\par

\section{Conditional ``amplitude'' operator}

Let us a consider a bipartite system $AB$, characterized by a density
operator $\rho_{AB}$ in the product Hilbert space 
${\cal H}_{AB}={\cal H}_A \otimes {\cal H}_B$. Each subsystem,
$A$ or $B$, is characterized by the reduced density operator
$\rho_A = {\rm Tr}_B [\rho_{AB}]$ or
$\rho_B = {\rm Tr}_A [\rho_{AB}]$, respectively.

\medskip
\noindent {\bf Definition 1:} Define the conditional amplitude
operator of $A$ conditional on $B$ 
as~\cite{bib_neginfo,bib_oviedo,bib_physcomp} 
\begin{eqnarray}  \label{eq_def_rhoA|B}
\rho_{A|B} &\equiv& \exp [ \log \rho_{AB} - 
\log ({\bf 1}_A \otimes \rho_B ) ]  \nonumber \\
&=& \lim_{n\to \infty} [\rho_{AB}^{1/n} 
({\bf 1}_A \otimes \rho_B )^{-1/n} ]^n
\end{eqnarray}
which is a positive semi-definite Hermitian operator in the joint
Hilbert space ${\cal H}_{AB}$ 
defined on the range of $\rho_{AB}$ (see Lemma 1). 
\par
\medskip

The second expression in Eq.~(\ref{eq_def_rhoA|B}), 
based on the Trotter decomposition 
of $\rho_{A|B}$, explicitly emphasizes that
the conditional amplitude operator
is the natural quantum analog of the conditional probability,
$p(a|b)=p(a,b)/p(b)$. As $\rho_{AB}$ and $({\bf 1}_A \otimes \rho_B)^{-1}$
do not necessarily commute, the Trotter symmetrization
is used here to obtain an Hermitian operator.

\medskip
\noindent {\bf Lemma 1:} 
${\rm Ran}(\rho_{AB})\subseteq {\rm Ran}({\bf 1}_A \otimes \rho_B)$, 
where ${\rm Ran}(\rho)$ is the range of $\rho$.\footnote{For a linear
and Hermitian operator $\rho$, the range is the subspace of the domain
of $\rho$ that is spanned by the eigenvectors corresponding to
non-zero eigenvalues, i.e., it is the support of $\rho$.}
Thus, the support of $\rho_{AB}$ is included in the
support of ${\bf 1}_A \otimes \rho_B$, and, consequently,
the conditional amplitude operator $\rho_{A|B}$
is well-defined on the support of $\rho_{AB}$.
\par
\medskip
We must prove that 
${\rm Ker}({\bf 1}_A \otimes \rho_B)\subseteq {\rm Ker}(\rho_{AB})$,
where ${\rm Ker}(\rho)$ is the kernel of $\rho$, i.e.,
that any eigenvector $|\psi\rangle$ of 
$({\bf 1}_A \otimes \rho_B)$ with zero eigenvalue is such that
$\rho_{AB}|\psi\rangle =0$. First note that 
any such eigenvector $|\psi\rangle$ can be written as a linear
combination of states $|\phi\rangle$ where
\begin{equation}
|\phi\rangle = |a\rangle \otimes |b\rangle
\end{equation}
and $|a\rangle$ is an arbitrary state vector in ${\cal H}_A$
while $|b\rangle$ is an eigenvector of $\rho_B$ with zero eigenvalue, i.e.,
$\rho_B|b\rangle=0$.
Let us now consider the positive semi-definite operator 
${\tilde \rho} \equiv
({\bf 1}_A \otimes P_b) \rho_{AB} ({\bf 1}_A \otimes P_b)$,
with $P_b = |b\rangle\langle b|$. It is trivial to check that
its partial trace over $A$ vanishes, that is
\begin{equation}
{\rm Tr}_A [{\tilde \rho}] = P_b \rho_B P_b = 0 
\end{equation}
This results from the general relation
\begin{equation}  \label{eq_useful}
{\rm Tr}_A [({\bf 1}_A \otimes \lambda_B) \mu_{AB}]
= \lambda_B {\rm Tr}_A [\mu_{AB}]
\end{equation}
where $\lambda_B$ and $\mu_{AB}$ are arbitrary operators in
${\cal H}_B$ and  ${\cal H}_{AB}$ respectively.
Since ${\tilde \rho}$ is positive semi-definite and traceless,
we have ${\tilde \rho}=0$. Thus, in particular,
the expectation value of ${\tilde \rho}$ in the state $|\phi\rangle$ vanishes,
\begin{equation}
\langle \phi | {\tilde \rho} | \phi \rangle =
\langle \phi | \rho_{AB} | \phi \rangle = 0
\end{equation}
which in turn implies that $\rho_{AB}|\phi\rangle =0 $ since $\rho_{AB}$
is positive semi-definite. As this is true for each term
$|\phi\rangle$ in the superposition, we conclude that 
$\rho_{AB}|\psi\rangle =0 $. $\Box$
\par

\medskip
\noindent {\bf Remark:}
Lemma 1 clearly implies that 
${\rm Ker}({\bf 1}_A \otimes \rho_B)\cap {\rm Ran}(\rho_{AB})= \emptyset$. 
Thus, the subspace spanned by the eigenvectors
with zero eigenvalue of $({\bf 1}_A \otimes \rho_B)$ is disjoint
from the support of $\rho_{AB}$, so that the definition of $\rho_{A|B}$
contains no singularities in the support of $\rho_{AB}$.
Of course, there is a classical analog for probability distributions
which ensures that $p(a|b)=p(a,b)/p(b)$ is
well defined if $a,b$ are such that $p(a,b)\ne 0$. Indeed, if $b$ is
such that $p(b)=0$, then $p(a,b)=0$, $\forall a$. This is obvious
since $p(b)=\sum_a p(a,b)$ and $p(a,b)\ge 0$.

\medskip
\noindent {\bf Definition 2:} The conditional von Neumann entropy
is defined using the joint density operator $\rho_{AB}$ and the conditional
amplitude operator $\rho_{A|B}$ as~\cite{bib_neginfo,bib_oviedo,bib_physcomp} 
\begin{equation}  \label{eq_def_S_A|B}
S(A|B) = - {\rm Tr}[ \rho_{AB} \log_2 \rho_{A|B} ]
\end{equation}
in close analogy to the classical definition
\begin{equation}
H(A|B) = - \sum_{a,b} p(a,b) \log_2 p(a|b)
\end{equation}
Thus, $S(A|B)$ corresponds to the {\em quantum} entropy of $A$
conditional on $B$, and is mathematically well-defined
as a consequence of Lemma 1. The trace in Eq.~(\ref{eq_def_S_A|B})
can indeed be restricted to the support of $\rho_{AB}$, using the fact
that the common eigenvectors $|\psi\rangle$ with zero eigenvalue 
of both $\rho_{AB}$ and $({\bf 1}_A \otimes \rho_B)$ 
yield a vanishing contribution to the
entropy, as $\lim_{x\to 0} (x \log x) =0$. (The same argument is used
in classical information theory to discard zero probabilities when 
calculating Shannon entropies.)
\par

\medskip
\noindent {\bf Theorem 1:} The definitions of $\rho_{A|B}$ and
the conditional von Neumann entropy
imply that $S(A|B)=S(AB)-S(B)$, as for Shannon entropies.
\par
\medskip
First, using Eqs.~(\ref{eq_def_rhoA|B}) and (\ref{eq_def_S_A|B}), we have
\begin{equation}  \label{eq_2terms}
S(A|B) = - {\rm Tr}[ \rho_{AB} \log_2 \rho_{AB} ]
+ {\rm Tr}[ \rho_{AB} \log_2 ({\bf 1}_A \otimes \rho_B ) ]
\end{equation}
where the first term on the right-hand side is clearly equal to $S(AB)$.
In order to calculate the second term on the right-hand side
of Eq.~(\ref{eq_2terms}), we write
\begin{eqnarray}
{\rm Tr}_A [ \rho_{AB} \log_2 ({\bf 1}_A \otimes \rho_B ) ]
&=& {\rm Tr}_A [ \rho_{AB} ({\bf 1}_A \otimes \log_2 \rho_B ) ] \nonumber\\
&=& {\rm Tr}_A [\rho_{AB}] \log_2 \rho_B \nonumber\\
&=& \rho_B \log_2 \rho_B
\end{eqnarray}
where we have made use of Eq.~(\ref{eq_useful}).
This implies that the second term on the right-hand side of 
Eq.~(\ref{eq_2terms}) is
\begin{equation}
{\rm Tr}_B [\rho_B \log_2 \rho_B] = - S(B)
\end{equation}
resulting in $S(A|B)=S(AB)-S(B)$. $\Box$
\par

\medskip
\noindent {\bf Lemma 2:} The spectrum of the conditional
amplitude operator $\rho_{A|B}$ is invariant under unitary
transformations of the product form $U_A \otimes U_B$ on $\rho_{AB}$.
\par

\medskip
Let us consider the isomorphism
\begin{equation}
\rho_{AB} \to \rho_{AB}'=(U_A \otimes U_B)\rho_{AB}
(U_A^{\dagger} \otimes U_B^{\dagger})
\end{equation}
We first calculate the partial trace of the joint density operator over $A$ 
{\em after} this transformation, that is
\begin{eqnarray}   \label{eq_lemma2eq}
\rho_B' &=& {\rm Tr}_A[\rho_{AB}'] \nonumber\\
&=& {\rm Tr}_A[ (U_A \otimes U_B)\rho_{AB}
(U_A^{\dagger} \otimes U_B^{\dagger})]  \nonumber \\
&=& {\rm Tr}_A[ ({\bf 1}_A \otimes U_B) (U_A \otimes {\bf 1}_B) \rho_{AB}
(U_A^{\dagger} \otimes {\bf 1}_B) ({\bf 1}_A \otimes U_B^{\dagger})]
\nonumber \\
&=& U_B {\rm Tr}_A[ (U_A \otimes {\bf 1}_B) \rho_{AB}
(U_A^{\dagger} \otimes {\bf 1}_B) ] U_B^{\dagger}
\nonumber \\
&=& U_B \rho_B U_B^{\dagger}
\end{eqnarray}
where we have used Eq.~(\ref{eq_useful})
and the basis invariance of the trace. 
This implies that the conditional amplitude operator
transforms as
\begin{equation}  \label{eq_iso_cond}
\rho_{A|B} \to \rho_{A|B}'=(U_A \otimes U_B)\rho_{A|B}
(U_A^{\dagger} \otimes U_B^{\dagger})
\end{equation}
so that its spectrum is conserved under $U_A \otimes U_B$ on
$\rho_{AB}$. Note that the classical analog of an $U_A\otimes U_B$
isomorphism corresponds to permuting the rows and columns
of the joint probability distribution $p(a,b)$, so that the
classical counterpart of Eq.~(\ref{eq_iso_cond}) is straightforward.
$\Box$
\par

\medskip
\noindent {\bf Remark:}
This Lemma suggests that the spectrum of $\rho_{A|B}$ could 
be related to the separability of the state $\rho_{AB}$, since separability
(or inseparability) is conserved under a $U_A \otimes U_B$
isomorphism. This will be examined later on.
\par

\medskip
\noindent {\bf Corollary:} The conditional von Neumann entropy $S(A|B)$ is
invariant under a unitary
transformation of the product form $U_A \otimes U_B$. 
\par
\medskip

This property 
results from the definition of $S(A|B)$, Eq.~(\ref{eq_def_S_A|B}), 
together with Eq.~(\ref{eq_iso_cond}), or can be
checked trivially from Theorem~1.
\par

\medskip
\noindent
{\bf Theorem 2:} The operator $\sigma_{AB} \equiv - \log \rho_{A|B} =
\log ({\bf 1}_A \otimes \rho_B ) - \log \rho_{AB}$ 
is positive semi-definite if the quantum bipartite system
characterized by $\rho_{AB}$ is separable.
\par

\medskip 
Let us consider
a separable bipartite system $\rho_{AB}$, i.e., a convex combination
of product states:
\begin{equation}  \label{eq_separablestate}
\rho_{AB} = \sum_i w_i \left( \rho_A^{(i)} \otimes \rho_B^{(i)} \right)
\qquad {\rm with~}\sum_i w_i=1 {\rm ~and~}0\le w_i \le 1
\end{equation}
where $\rho_A^{(i)}$ and $\rho_B^{(i)}$ are states in ${\cal H}_A$ and
${\cal H}_B$, respectively. We first define the operator
\begin{equation}
\lambda_{AB} \equiv ({\bf 1}_A \otimes \rho_B ) - \rho_{AB}
\end{equation}
It is easy to check that
$\lambda_{AB}$ is positive semi-definite if $\rho_{AB}$ is
separable. Indeed, in such a case we have
\begin{equation}   \label{eq_separa}
\lambda_{AB} = \sum_i w_i 
\Big( \underbrace{( {\bf 1}_A-\rho_A^{(i)})}_{\ge 0}
\otimes \underbrace{\rho_B^{(i)}}_{\ge 0} \Big)  \ge 0
\end{equation}
since a sum of positive operators is a positive operator.
Now, we can use the fact that, if $X$ and $Y$ are two Hermitian operators
such that $X\ge Y > 0$,\footnote{Here and below, the notation $X \ge Y$
means that $X-Y$ is a positive semi-definite operator.}
then $\log X \ge \log Y$, as implied by
L\"owner's theorem~\cite{bib_horn}. (Note that the converse is not true.)
As a consequence, using $X={\bf 1}_A \otimes \rho_B $
and $Y=\rho_{AB}$, we conclude that 
$\lambda_{AB} \ge 0$ implies $\sigma_{AB} \ge 0$. $\Box$
\par

\medskip
\noindent {\bf Corollary 1:} Any separable bipartite state is such that
$\rho_{A|B} \le 1$.
\par
\medskip

Since we have $\rho_{A|B}=\exp(-\sigma_{AB})$, Theorem 2 shows indeed that 
no eigenvalue of the conditional amplitude
operator exceeds 1 for a separable state, as was conjectured 
in Ref.~\cite{bib_neginfo,bib_oviedo,bib_physcomp}. 
This yields a simple {\em necessary} (but not sufficient)
condition for separability. The classical analog of this property is that
$-\log p(a|b) \ge 0$, $\forall a,b$. The latter inequality simply results
from the fact that $p(a|b)=p(a,b)/p(b)\le 1$, $\forall a,b$, as 
$p(b)=\sum_a p(a,b)$ and $p(a,b)\ge 0$.

\medskip
\noindent {\bf Corollary 2:} The conditional von Neumann entropy
$S(A|B)$ is non-negative for a separable bipartite state.
\par
\medskip

Since we have $S(A|B)= {\rm Tr}[\rho_{AB} \; \sigma_{AB}]$, this
simply follows from the fact that ${\rm Tr}[XY]\ge 0$ if
$X,Y\ge 0$. Thus, the non-negativity of the conditional
entropy is another (weaker) necessary condition
for separability~\cite{bib_neginfo,bib_oviedo,bib_physcomp}.
This has been shown in general for ``$\alpha$-entropies''
in Ref.~\cite{bib_hor_alphaentropy}. It can also be related to
the non-violation of entropic Bell inequalities~\cite{bib_bell}.
\par

Note that Corollary 2 can also be obtained by using the concavity
of $S(A|B)$ in a convex combination of $\rho_{AB}$'s, 
\begin{equation}
S(A|B)=S(\rho_{AB})-S(\rho_B) \ge \sum_i w_i 
\left( S(\rho_{AB}^{(i)})-S(\rho_B^{(i)}) \right)
\qquad {\rm if~}
\rho_{AB}=\sum_i w_i \rho_{AB}^{(i)}
\end{equation}
a property related to the strong subadditivity of 
quantum entropies~\cite{bib_wehrl}.
Using the fact that, for a separable state, each term $i$ gives
$S(A|B)=S(A)$ because $\rho_{AB}^{(i)}=\rho_A^{(i)} \otimes \rho_B^{(i)}$
(i.e., $A$ and $B$ are independent), we obtain
\begin{equation}
S(A|B) \ge \sum_i w_i S(\rho_A^{(i)}) \ge 0
\end{equation}
Note that a negative conditional von Neumann entropy $S(A|B)$
necessarily implies that an eigenvalue of $\rho_{A|B}$ exceeds 1, 
but the converse is not true. Thus, weak inseparability (in the sense
that $S(A|B)\ge 0$ despite the inseparability of $\rho_{AB}$)
may be revealed by the spectrum of $\rho_{A|B}$.
\par

\medskip
\noindent
{\bf Theorem 3:} There exist inseparable bipartite states $\rho_{AB}$
such that the operator $\sigma_{AB}$ is positive semi-definite;
consequently, $\sigma_{AB}\ge 0$ (or $\rho_{A|B} \le 1$)
is {\em not} a sufficient condition for separability.
\par

\medskip
Let us consider a bipartite system $AB$ characterized by $\rho_{AB}$,
which we extend with another system $A'B'$ in the state $\rho_{A'B'}$.
The joint system is then characterized by a density operator
of the product form
\begin{equation}
\rho_{AA';BB'}= \rho_{AB} \otimes \rho_{A'B'}
\end{equation}
We first calculate the conditional amplitude operator of the
joint system ($AA'$ conditional on $BB'$)
\begin{equation}
\rho_{AA'|BB'}= \exp [ \log \rho_{AA';BB'} - 
\log ({\bf 1}_{AA'} \otimes \rho_{BB'} ) ] 
\end{equation}
where the reduced density operator describing $BB'$ is
\begin{equation}
\rho_{BB'} = {\rm Tr}_{AA'} [\rho_{AA';BB'}] = \rho_B \otimes \rho_{B'} 
\end{equation}
Using the identity
$\log(X \otimes Y) = \log X \otimes 1 + 1 \otimes \log Y $
for operators $X,Y>0$ as well as its exponential, i.e.,
$\exp X \otimes \exp Y = \exp (X \otimes 1 + 1 \otimes Y)$, 
we obtain
\begin{eqnarray}
\rho_{AA'|BB'} &=& \exp [ \log \rho_{AB} \otimes {\bf 1}_{A'B'}
                         + {\bf 1}_{AB} \otimes \log \rho_{A'B'}
- {\bf 1}_A \otimes \log \rho_B \otimes {\bf 1}_{A'B'}
- {\bf 1}_{AB} \otimes {\bf 1}_{A'} \otimes \log \rho_{B'} ]
\nonumber \\
&=& \exp [ (\log \rho_{AB}- {\bf 1}_A \otimes \log \rho_B )
       \otimes {\bf 1}_{A'B'}  + {\bf 1}_{AB} 
       \otimes (\log \rho_{A'B'} - {\bf 1}_{A'} \otimes \log \rho_{B'}) ]
\nonumber \\
&=& \exp [\log \rho_{AB}- {\bf 1}_A \otimes \log \rho_B ] \otimes
    \exp [\log \rho_{A'B'} - {\bf 1}_{A'} \otimes \log \rho_{B'} ]
\end{eqnarray}
Thus, we have
\begin{equation}  \label{eq_productcond}
\rho_{AA'|BB'} = \rho_{A|B} \otimes \rho_{A'|B'}
\end{equation}
which parallels the classical relation $p(aa'|bb')=p(a|b) p(a'|b')$ 
if $AB$ and $A'B'$ are independent bipartite systems, that is,
if $p(a,a';b,b')=p(a,b) p(a',b')$. In particular, we have
\begin{equation}   \label{eq_productcondspectrum}
\sigma(\rho_{AA'|BB'})=\sigma(\rho_{A|B})\otimes \sigma(\rho_{A'|B'})
\end{equation}
where $\sigma(\rho)$ stands for the spectrum of $\rho$.
\par

Now, let us assume that $AB$ is an inseparable system 
with $\sigma_{AB} \not\ge 0$ or $\rho_{A|B} \not\le 1$. In other
words, the operator $\rho_{A|B}$ admits an eigenvalue that exceeds 1.
Assume also that $A'B'$ corresponds to two independent systems
in a product state, that is, $\rho_{A'B'}=\rho_{A'} \otimes \rho_{B'}$. 
The resulting conditional amplitude operator for $A'B'$ is 
then $\rho_{A'|B'}=\rho_{A'}\otimes {\bf 1}_{B'}$,
just like its classical counterpart $p(a|b)=p(a)$ if
$p(a,b)=p(a)p(b)$.  Obviously, we then
have $\rho_{A'|B'}\le 1$, as expected since $A'B'$ is separable.
According to Eq.~(\ref{eq_productcondspectrum}),
the eigenvalues of $\rho_{AA'|BB'}$ are the pairwise products of
eigenvalues of $\rho_{A|B}$ with eigenvalues of $\rho_{A'|B'}$.
Therefore, it is easy to find a system $A'B'$ with eigenvalues of
$\rho_{A'|B'}$ small enough so that the product of any of them
with an unclassical ($>1$) eigenvalue of $\rho_{A|B}$ results in eigenvalues
of $\rho_{AA'|BB'}$ that are all $\le 1$. The extended system
is then characterized by $\sigma_{AA';BB'} \ge 0$ or
$\rho_{AA'|BB'}\le 1$, while it obviously contains 
an inseparable component $AB$.
Such a {\em dilution} of inseparability (or entanglement)
is always achievable
with a system $A'B'$ that is large enough and maximally disordered
(i.e., $\rho_{A'B'} \sim {\bf 1}_{A'} \otimes {\bf 1}_{B'}$).
Consequently, the condition that $\sigma_{AB} \ge 0$ or $\rho_{A|B}\le 1$
cannot be a sufficient condition for separability. 
$\Box$
\par

\medskip
\noindent {\bf Remark 1:}
Eq.~(\ref{eq_productcondspectrum}) implies that, if $AB$ and $A'B'$
are inseparable systems with $\rho_{A|B}\not\le 1$ and 
$\rho_{A'|B'}\not\le 1$, then the inseparability of the joint system
is revealed by $\rho_{AA'|BB'}\not\le 1$.
\par

\medskip
\noindent {\bf Remark 2:}
While $\lambda_{AB}\ge 0$ is
a sufficient separability condition for $2\times 2$ and
$2\times 3$ systems (cf. Section III), it cannot be concluded 
that $\sigma_{AB} \ge 0$ or 
$\rho_{A|B}\le 1$ is also sufficient in these cases, as the converse of 
L\"owner's theorem does not hold. Interestingly, 
numerical evidence reveals that only very few inseparable states
of two qubits with $\rho_{A|B}\le 1$ can be found.
\par

\section{Necessary condition for separability of mixed states}

\subsection{Bipartite system of arbitrary dimension}

As we saw in the previous section, Theorem 2
results in a simple {\it necessary} condition for the separability
of mixed states based on Eq.~(\ref{eq_separa}), which does not require
the calculation of the conditional amplitude operator (although it
is related to it).
\par
\medskip
\noindent {\bf Definition 3:}
Define a linear map $\Lambda$ which maps
Hermitian operators on ${\cal H}_{AB}$ into 
Hermitian operators on ${\cal H}_{AB}$:
\begin{equation}
\Lambda : \rho_{AB} \to \lambda_{AB} \equiv
{\bf 1}_A \otimes \rho_B  - \rho_{AB}
\qquad~{\rm with~}\rho_B={\rm Tr}_A [\rho_{AB}]
\end{equation}
It commutes with a unitary transformation acting
independently on $A$ and $B$. Indeed, according to Eq.~(\ref{eq_lemma2eq}),
if $\rho_{AB}$ undergoes a unitary transformation
of the product form $U_A \otimes U_B$, then 
\begin{equation}
\lambda_{AB} \to \lambda_{AB}'=(U_A \otimes U_B)\lambda_{AB}
(U_A^{\dagger} \otimes U_B^{\dagger})
\end{equation}
i.e., $\lambda_{AB}$ transforms just like $\rho_{AB}$.
Therefore, the spectrum of $\lambda_{AB}$ is invariant under
a $U_A \otimes U_B$ isomorphism on $\rho_{AB}$, as expected.

\medskip
\noindent
{\bf Theorem 4:} A necessary condition for the separability of the
state $\rho_{AB}$ of a bipartite system $AB$ is that it is mapped
by $\Lambda$ into a non-negative operator, 
i.e., $\Lambda \rho_{AB} \ge 0$.
\par
\medskip

The proof of this theorem is contained in the 
proof of Theorem 2 where we used the fact that
the map $\Lambda$ reveals non-separability: if $\lambda_{AB} \not\ge 0$, then
$\rho_{AB}$ is inseparable. Conversely, any separable state $\rho_{AB}$
is mapped into a positive semi-definite operator $\lambda_{AB}$.
Moreover, it is easy to see that
the map $\Lambda$ conserves separability since it is linear
and maps product states into product states: if $\rho_{AB}$ is
separable, then $\lambda_{AB}\ge 0$ is also separable. 
Let us now calculate the partial traces of the operator $\lambda_{AB}$:
\begin{eqnarray}
\lambda_A &=& {\rm Tr}_B [\lambda_{AB}] = {\bf 1}_A - \rho_A \\
\lambda_B &=& {\rm Tr}_A [\lambda_{AB}] = (d_A-1) \rho_B
\end{eqnarray}
where $d_A$ is the dimension of ${\cal H}_A$.
This shows that the map $\Lambda$ does not
preserve the trace in general. Indeed, 
the total trace is scaled by an integer factor under $\Lambda$, that is,
${\rm Tr}[\lambda_{AB}] = (d_A-1) {\rm Tr}[\rho_{AB}]$.
Thus, $\Lambda$ is {\em trace-preserving} only
in the special case where $A$ is a 2-state system (i.e., $d_A=2$).
It is also interesting to note that the map $\Lambda$ is 
always {\em reversible}, the inverse map being given by
\begin{equation}
\Lambda^{-1}: \lambda_{AB} \to (d_A -1)^{-1} ({\bf 1}_A \otimes \lambda_B) 
 - \lambda_{AB} = \rho_{AB}
\end{equation}
where $\lambda_B$ is defined as above.
Notice that the map $\Lambda$ is equal to its inverse $\Lambda^{-1}$
only in the case where $d_A=2$. 
This implies that, if $\lambda_{AB}$ is separable,
then $\Lambda^{-1}:\lambda_{AB}\to\rho_{AB}\ge 0$. (The fact that the
inverse map reveals inseparability is true in this case only.) 
\par

The separability condition based on Theorem 4 is illustrated in
Appendix~A, where we consider several separable and inseparable
states. It appears that $\lambda_{AB}\ge 0$ results in the same condition
as Peres' in the case of two quantum bits, in which case it is sufficient
(see Theorem 9). For larger dimensions, it is only necessary.
\par

\medskip
\noindent {\bf Remark 1:}
More generally, following the approach of Horodecki et
al.~\cite{bib_horo_sufficient}, the map $\Lambda$ can be written
as the tensor product of a positive linear map $\Gamma$ and the identity,
that is
\begin{equation}
\Lambda= \Gamma \otimes I 
\qquad {\rm with~}\Gamma: \rho \to ({\rm Tr}\rho)-\rho
\end{equation}
where $\Gamma$ acts on Hermitian operators on ${\cal H}_A$ and the identity
acts on operators on ${\cal H}_B$. Since $\Gamma$ is a positive map
(i.e., it maps positive operators into positive operators), 
$\Lambda=\Gamma \otimes I$ maps {\em separable} states 
into {\em positive} operators~\cite{bib_horo_sufficient}. 
It therefore results in a
{\em necessary} condition for separability, according to Theorem 4.
The map $\Gamma$ commutes with an arbitrary 
unitary transformation $U$, that is
\begin{equation}
\Gamma (U \rho U^{\dagger}) = U (\Gamma \rho) U^{\dagger}
\end{equation}
which makes the separability condition based 
on $\Lambda=\Gamma\otimes I$
independent on the basis chosen for $A$ and $B$.
In the same manner, the inverse map $\Lambda^{-1}$ can be written as
\begin{equation}
\Lambda^{-1}= \Gamma^{-1} \otimes I 
\qquad {\rm with~}\Gamma^{-1}: \rho \to {{\rm Tr}\rho \over d-1} -\rho
\end{equation}
where $d$ is the dimension of the Hilbert space of $\rho$.
Note that $\Gamma^{-1}$ is {\em not} a positive map for $d>2$, so that
$\Lambda^{-1}$ is in general useless as far as 
detecting inseparability is concerned.
This emphasizes that the separability criterion based on Theorem~4
is quite special in 2 dimensions (e.g., for a spin-1/2 particle or a
quantum bit), as will be studied in detail
later on. Let us only mention here that the map $\Gamma$ applied to
a two-dimensional system amounts to a spin flip
(this can be interpreted as time reversal). In order to see this, 
let us write an arbitrary state of a two-dimensional system as
\begin{equation}  \label{eq_bloch}
\rho={1\over 2} (1+\vec{r}\cdot \vec{\sigma})
\end{equation}
where $\vec{\sigma}$ represent the three Pauli matrices and 
$\vec{r}={\rm Tr}(\rho \vec{\sigma})$ is
a {\em real} vector in the Bloch sphere (of radius 1). The vector
$\vec{r}$ describes the statistics of measurements on the system, as,
for example, the quantum expectation value of the
spin component along an axis defined by the vector $\vec{v}$ is 
\begin{equation}
{\rm Tr}\Big[\rho (\vec{v}\cdot \vec{\sigma})\Big]=(\vec{v}, \vec{r})
\end{equation}
Using Eq.~(\ref{eq_bloch}), it is straightforward to check that
\begin{equation}  \label{eq_gammarho}
\Gamma \rho = 1 - \rho 
            = {1\over 2} (1-\vec{r}\cdot \vec{\sigma})
\end{equation}
Thus, the map $\Gamma$ performs a spin-flip, or, equivalently,
a parity transformation $\vec{r} \to - \vec{r}$
on the vector characterizing the system in the Bloch sphere.
\par

\medskip
\noindent {\bf Remark 2:}
It is interesting to consider the classical analog of the maps
$\Gamma$ and $\Lambda=\Gamma\otimes I$ to gain some insight into
their physical meaning.
First, the map $\Gamma$ applied to a classical probability
distribution $p_i$ (diagonal $\rho$) corresponds to the transformation:
\begin{equation}
p_i \to q_i = \sum_k p_k - p_i
\end{equation}
(Obviously, $q_j \ge 0$ is not normalized except for a binary
distribution.) The classical analog of the map $\Lambda = \Gamma \otimes I$
is thus
\begin{equation}
p_{ij} \to q_{ij}=\left( \sum_k p_{k|j} - p_{i|j}\right) p_j = p_j - p_{ij}
\end{equation}
Since $p_{i|j}$ is a probability distribution in $i$, we always have
$1-p_{i|j} \ge 0$ so that $q_{ij}\ge 0$ and the separability
criterion is fulfilled. This emphasizes that quantum
inseparability (``$q_{ij} <0$'') may be viewed as resulting from 
a conditional probability that {\em exceeds} 1 (more precisely, an eigenvalue 
of $\rho_{A|B}$ which exceeds 1).
\par

\medskip
\noindent {\bf Remark 3:}
It is instructive to consider the action of the map $\Lambda=\Gamma\otimes I$
on a product state $|\psi\rangle = |a\rangle \otimes |b\rangle$. Using
$\rho_{AB}=P_a \otimes P_b$ with $P_a=|a\rangle \langle a|$
and $P_b=|b\rangle \langle b|$,
it is easy to see that
\begin{equation}
\lambda_{AB}=P_a^{\perp}\otimes P_b
\end{equation}
where $P_a^{\perp}= \Gamma (|a\rangle \langle a|) =
{\bf 1}_A - |a\rangle \langle a|$
is the projector on the subspace orthogonal to $|a\rangle$.
In the special case where $A$ is a 2-state system ($d_A=2$),
$P_a^{\perp}$ is a rank-one projector as the total trace is preserved.
We have then $P_a^{\perp}=|a^{\perp}\rangle \langle a^{\perp}|$,
where $|a^{\perp}\rangle$ is a state vector 
orthogonal to $|a\rangle$ obtained by applying
a complex conjugation on the components of $|a\rangle$ followed
by a rotation by an angle $\pi$ about the $y$-axis. (Note that it
can be shown that it is impossible to construct a state $|a^{\perp}\rangle$
that is orthogonal to an {\em arbitrary} state $|a\rangle$ by applying
a unitary transformation alone.)
Indeed, an arbitrary state
\begin{equation}
|a\rangle= \alpha|0\rangle +\beta|1\rangle
\end{equation} 
with $|\alpha|^2 +|\beta|^2=1$, is transformed into
\begin{equation}
|a^{\perp}\rangle= -\beta^*|0\rangle+\alpha^*|1\rangle
\end{equation}
by applying to the
state vector $\alpha^*|0\rangle+\beta^*|1\rangle$ the rotation
\begin{equation}
U_y=\exp(-i\pi\sigma_y/2)=-i\sigma_y=\sigma_x\sigma_z= 
\left( \begin{array}{cc}
 0 & -1 \\
 1 & 0  \end{array} \right)
\end{equation}
that is, a bit- and phase-flip.
The transformed state $|a^{\perp}\rangle$ is such that
$\langle a^{\perp} | a \rangle =0$ and 
$|a^{\perp}\rangle \langle a^{\perp}|= {\bf 1}_A - |a\rangle \langle a|$,
as expected.
Consequently, the $\Gamma$ map (i.e., the complex conjugation followed
by $U_y$ rotation) is an {\em antiunitary} operation on state vectors
in the 2-dimensional Hilbert space ${\cal H}_A$. Indeed,
for any two state vectors $|a \rangle$ and $|{\tilde a} \rangle$, we have
$\langle {\tilde a}^{\perp}|a^{\perp}\rangle = \langle {\tilde a}|a\rangle^*$.
Since time-reversal $T$ is known to be an antiunitary operator given by
complex conjugation $K$ followed by 
a rotation ${\cal R}_y$ of angle $\pi$ about the $y$-axis,
i.e., $T={\cal R}_y K$ (see, e.g., \cite{bib_schiff}),
we conclude that the map $\Gamma$ is time-reversal when $d_A=2$. 
[This is also clear from Eq.~(\ref{eq_gammarho}).]
Consequently, when $d_A=2$, $\Lambda=\Gamma\otimes I$
corresponds to physical time-reversal on the
subsystem $A$ (while leaving the subsystem $B$ unchanged). Such a link
between ``local'' time-reversal $T\otimes I$ and separability
has recently been pointed out by Sanpera et al.~\cite{bib_sanpera}.
The discussion of the case $d_A=d_B=2$ will be continued in Section IIIb.

\medskip
\noindent {\bf Lemma 3:}
If $\rho_{AB}$ is a separable state, then $\lambda_{AB}$
is a separable operator obtained by replacing the states $|a\rangle$
in ${\cal H}_A$ by projectors $P_a^{\perp}$ orthogonal to them.
\par
\medskip

Let us assume that the bipartite system $AB$ is separable, that is
\begin{equation}   \label{eq_sep_ab}
\rho_{AB} = \sum_i w_i \left( |a_i\rangle \langle a_i|
            \otimes |b_i\rangle \langle b_i| \right)
\end{equation}
where the $|a_i\rangle \otimes |b_i\rangle$ are pure product
states [using the spectral decomposition of $\rho_A^{(i)}$ 
and $\rho_B^{(i)}$, it is easy to rewrite
Eq.~(\ref{eq_separablestate}) into this form].
As a result of Remark 3, we see that $\rho_{AB}$
it is mapped by $\Lambda$ into a separable operator of the form
\begin{equation}   
\lambda_{AB} = \sum_i w_i \left( P_{a_i}^{\perp} \otimes 
    |b_i\rangle \langle b_i|  \right)  \qquad \Box
\end{equation}
The operator $\lambda_{AB}$ is a unit-trace operator
in the case $d_A=2$ since each component pure state
$|a\rangle \otimes |b\rangle$ is mapped into
a {\em pure} product state, $|a^{\perp}\rangle \otimes |b\rangle$,
in which case it simply reads
\begin{equation}  \label{eq_sep_aperpb}
\lambda_{AB} = \sum_i w_i \left( |a_i^{\perp}\rangle \langle a_i^{\perp}|
\otimes |b_i\rangle \langle b_i|  \right)
\end{equation}
This expression results in a simple {\em necessary} condition for separability
(distinct from $\lambda_{AB}\ge0$), inspired from the condition
recently proposed by Horodecki~\cite{bib_horo_one} as follows.
\par

\medskip
\noindent
{\bf Theorem 5:} A necessary separability condition for the
bipartite state $\rho_{AB}$ is that its support can be spanned by
a set of product states which are such that the corresponding
product states obtained by applying $\Gamma$ to the state vector
in the $A$ space span the support of $\lambda_{AB}= \Lambda \rho_{AB}$.
\par
\medskip

We only consider this condition in the case where $d_A=2$.
The central point is to note that, if $\rho_{AB}$ is separable
as in Eq.~(\ref{eq_sep_ab}), then the ensemble of product
states $|a_i\rangle \otimes |b_i\rangle$ span the entire
support of $\rho_{AB}$~\cite{bib_horo_one}. (Conversely, any state 
$|a_i\rangle \otimes |b_i\rangle$ must belong to the support of
$\rho_{AB}$ and cannot have a non-vanishing component orthogonal
to it.)
Using Lemma 3, we see that the ensemble
of states $|a_i^{\perp}\rangle \otimes |b_i\rangle$ span the entire
support of the corresponding separable state 
$\lambda_{AB}$ obtained by applying $\Lambda$ on $\rho_{AB}$
[cf. Eq.~(\ref{eq_sep_aperpb})]. (Also,  any state 
$|a_i^{\perp}\rangle \otimes |b_i\rangle$ cannot be outside the support
of $\lambda_{AB}$.) This results in another {\em necessary} condition for
separability which can be stated as follows. 
If a state $\rho_{AB}$ is separable, then it must be possible
to span its support by a set of product states $|a\rangle |b\rangle$
which are such that their image (i.e., the product states 
obtained by rotating the complex conjugate of state vector $|a\rangle$
in the $A$ space by an angle $\pi$ about the $y$-axis while leaving the 
state vector $|b\rangle$ in the $B$ space unchanged) span the support of
the mapped state $\lambda_{AB}=\Lambda \rho_{AB}$. 
$\Box$
\par

\medskip
\noindent
{\bf Definition 4:} Two additional maps from operators on ${\cal H}_{AB}$
to operators on ${\cal H}_{AB}$ can be defined: the {\em dual} map 
\begin{equation}  \label{eq_dualmap}
{\tilde \Lambda} : \rho_{AB} \to {\tilde \lambda}_{AB} 
= \rho_A \otimes {\bf 1}_B - \rho_{AB}
\end{equation}
and the {\em symmetric} map
\begin{equation}  \label{eq_symmetricmap}
M: \rho_{AB} \to \mu_{AB} = {\bf 1}_A \otimes {\bf 1}_B
- \rho_A \otimes {\bf 1}_B - {\bf 1}_A \otimes \rho_B + \rho_{AB}
\end{equation}
where $\rho_A={\rm Tr}_B [\rho_{AB}]$ and 
$\rho_B={\rm Tr}_A [\rho_{AB}]$
\par
\medskip

The map $\Lambda$ which we considered until now is related to the
conditional amplitude operator of $A$ conditionally on $B$, that
is $\rho_{A|B}$. Of course, a similar linear map can be defined 
using the amplitude operator $\rho_{B|A}$,
and exactly the same conclusions follow. This is the {\em dual} map 
${\tilde \Lambda}$ defined in Eq.~(\ref{eq_dualmap}).
It is trace-preserving and self-inverse in the special
case where $d_B=2$. It can obviously be written as the
tensor product ${\tilde \Lambda}= I \otimes \Gamma$, where
the map $\Gamma$ now acts on operators on ${\cal H}_B$, and
therefore commutes with a $U_A \otimes U_B$ isomorphism. Since
$\Gamma$ is positive, ${\tilde \Lambda}$ maps separable states
into positive (separable) operators, which results
in another separability condition, i.e., ${\tilde \lambda}_{AB}\ge 0$.
As we will see in Section IIIb, the operators $\lambda_{AB}$
and ${\tilde \lambda}_{AB}$ can be shown to have the same spectrum
when $d_A=d_B=2$ (i.e., for two 2-state systems), in which case
they result in the same separability condition. However, this
property does not hold in larger dimensions, i.e., $\lambda_{AB}$
and ${\tilde \lambda}_{AB}$ do not have the same spectrum in general
(see Appendix A).
\par

We can also construct another linear map 
by cascading $\Lambda$ and ${\tilde\Lambda}$ (the order is irrelevant),
which results in the {\em symmetric} map 
$M={\tilde \Lambda} \Lambda=\Gamma \otimes \Gamma$ defined
in Eq.~(\ref{eq_symmetricmap}). Any separable $\rho_{AB}$ is mapped
by $M$ into a separable operator $\mu_{AB}\ge 0$, as expected.
The symmetric map also commutes with a $U_A \otimes U_B$ isomorphism, 
\begin{equation}
M\left((U_A \otimes U_B) \rho_{AB}
      (U_A^{\dagger} \otimes U_B^{\dagger}) \right)=
(U_A \otimes U_B)(M\rho_{AB})(U_A^{\dagger} \otimes U_B^{\dagger})
\end{equation}
so that the spectrum of
$\mu_{AB}=M\rho_{AB}$ is invariant under local transformations on $\rho_{AB}$.
It is also reversible, its
inverse map $M^{-1}=\Gamma^{-1} \otimes \Gamma^{-1}$
being given by
\begin{equation}
M^{-1}: \mu_{AB} \to {\bf 1}_A \otimes {\bf 1}_B
- (d_B-1)^{-1} (\mu_A \otimes {\bf 1}_B) 
- (d_A-1)^{-1} ({\bf 1}_A \otimes \mu_B) + \mu_{AB}
  =\rho_{AB}
\end{equation}
where 
$\mu_A={\rm Tr}_B [\mu_{AB}]=(d_B-1)({\bf 1}_A - \rho_A)$,
$\mu_B={\rm Tr}_A [\mu_{AB}]=(d_A-1)({\bf 1}_B - \rho_B)$,
and $d_A$ ($d_B$) is the dimension of ${\cal H}_A$ (${\cal H}_B$).
As expected, this map is trace-preserving and self-inverse in the special
case where $d_A=d_B=2$. It corresponds then to a time-reversal
operation applied to the {\em joint} system $AB$. 
Note that, in this case, the symmetric map $M$ by itself is not
useful as far as revealing inseparability is concerned
since it is positive, i.e., $M\rho_{AB}\ge0$. 
Therefore, all inseparable states of two quantum bits
are mapped into positive operators $\mu_{AB}$ just as separable
states. Still, $M$ is important when analyzing the separability of
two quantum bits as it is also equivalent to the
conjugation operation in the ``magic'' basis introduced by
Hill and Wootters~\cite{bib_hillwootters} (see below, Theorem 10).
The case of two quantum bits will be studied later on. Whether
the positivity of $M$ holds in arbitrary dimensions is not known.
\par

\medskip
\noindent
{\bf Theorem 6:} The criterion that $\Lambda$ maps the state $\rho_{AB}$
into a positive semi-definite operator $\lambda_{AB}$ is {\em not}
a sufficient condition for the separability of $\rho_{AB}$.
\par
\medskip

The fact that $\lambda_{AB}\ge 0$ is necessary for
separability (Theorem 4) was proven before. In order to prove that 
it is not sufficient, we
will show that it is possible to have an {\em inseparable}
system with $\lambda_{AB} \ge 0$, i.e., such that its
inseparability is {\em not} revealed by the map $\Lambda$.
As before, we build an inseparable system $\rho_{AB}$ by extending an
inseparable component with a separable one.
Let us consider an inseparable system $A'B'$ with
$\lambda_{A'B'} \not\ge 0$.
We extend $A'B'$ with a separable system $A''B''$, and apply the
separability criterion to the joint system $AB$ where 
$A\equiv A'A''$ and $B\equiv B'B''$. Since the joint
system is characterized by the density operator 
$\rho_{AB}=\rho_{A'B'}\otimes \rho_{A''B''}$, 
its associated operator under the map $\Lambda$ is
given by
\begin{eqnarray}
\lambda_{AB} &=& \Lambda \rho_{AB}=
                 {\bf 1}_A \otimes \rho_B - \rho_{AB}  \nonumber\\
             &=& ({\bf 1}_{A'}\otimes \rho_{B'})
                 \otimes({\bf 1}_{A''}\otimes \rho_{B''})
                 - \rho_{A'B'}\otimes \rho_{A''B''}
\end{eqnarray}
Thus, using the operators 
$\lambda_{A'B'}=\Lambda \rho_{A'B'}=
{\bf 1}_{A'} \otimes \rho_{B'} - \rho_{A'B'} $ 
and $\lambda_{A''B''}=\Lambda \rho_{A''B''}=
{\bf 1}_{A''} \otimes \rho_{B''} - \rho_{A''B''}$
corresponding to $\Lambda$ applied to both component systems,
we obtain the simple expression
\begin{equation}   \label{eq_extendedlambda}
\lambda_{AB}=\lambda_{A'B'}\otimes \lambda_{A''B''} +
\lambda_{A'B'}\otimes \rho_{A''B''} + \rho_{A'B'}\otimes \lambda_{A''B''}
\end{equation}
with $\lambda_{A'B'} \not\ge 0$ and 
$\lambda_{A''B''} \ge 0$ (since $A''B''$ is separable). Thus,
the sum of the first two terms on the right-hand
side of Eq.~(\ref{eq_extendedlambda}) is {\em not} positive semi-definite,
while only the third one is. Therefore, Eq.~(\ref{eq_extendedlambda})
does {\em not} guarantee that $\lambda_{AB}\ge 0$ 
even though the composite system $AB$ contains an inseparable
component as $\lambda_{A'B'} \not\ge 0$. 
$\Box$
\par
\medskip

Note that, if both components are inseparable systems
that have $\lambda_{A'B'} \not\ge 0$
and $\lambda_{A''B''} \not\ge 0$, then $\lambda_{AB} \not\ge 0$
is not necessarily true, so that the inseparability of the joint
system $AB$ is not always revealed by the map $\Lambda$. [This
property contrasts with Eq.~(\ref{eq_productcond}).]
Conversely, Eq.~(\ref{eq_extendedlambda}) implies that,
if both components have $\lambda_{A'B'}\ge 0$
and $\lambda_{A''B''}\ge 0$, then $\lambda_{AB} \ge 0$.
The example of weakly inseparable states with a positive 
partial transpose (see Ref.~\cite{bib_horo_one}) is treated in Appendix A,
to illustrate that $\lambda_{AB} \ge 0$ is not a sufficient
condition in general.

\medskip
\noindent {\bf Remark:} The mechanism of dilution of inseparability
can be understood by examining the action of the map $\Gamma$ on
product states. Indeed, when applying the map $\Lambda=\Gamma \otimes I$
on the state $\rho_{AB}=\rho_{A'B'}\otimes \rho_{A''B''}$,
$\Gamma$ acts on the state $\rho_{A'}\otimes \rho_{A''}$
($B$ and $B'$ are left unchanged by $I$). Let us
consider a density operator
of the product form $\rho=\rho' \otimes \rho''$.
Since we have ${\rm Tr}(\rho)={\rm Tr}(\rho'){\rm Tr}(\rho'')$, 
we see that it is mapped to 
\begin{eqnarray}
\Gamma (\rho' \otimes \rho'')
&=& {\rm Tr}(\rho'){\rm Tr}(\rho'')- \rho' \otimes \rho''   \nonumber \\
&=& \left[ {\rm Tr}(\rho')-\rho' \right] \otimes 
    \left[ {\rm Tr}(\rho'')- \rho'' \right]
    +{\rm Tr}(\rho')\otimes \rho'' + \rho' \otimes {\rm Tr}(\rho'')
    - 2 \rho' \otimes \rho''  \nonumber \\
&=& \Gamma \rho' \otimes \Gamma \rho'' + \Gamma \rho' \otimes \rho''
    + \rho' \otimes \Gamma \rho''
\end{eqnarray}
which implies the relation
\begin{equation}
\Gamma=\Gamma' \otimes \Gamma'' + \Gamma' \otimes I'' + I' \otimes \Gamma''
\end{equation}
where $\Gamma'$ (or $\Gamma''$) stands for the same map but acting
on the subspace of $\rho'$ (or $\rho''$) while $\Gamma$ acts on the
joint space. Using the same notation for the map $\Lambda$ (i.e., 
$\Lambda'$ acts on the subspace of $A'B'$ while
$\Lambda''$ acts on the subspace of $A''B''$),
the latter equation gives
\begin{equation}  \label{eq_dilutionlambda}
\Lambda = \Gamma \otimes I = \Lambda' \otimes \Lambda'' + 
\Lambda' \otimes I'' + I' \otimes \Lambda''
\end{equation}
which results in Eq.~(\ref{eq_extendedlambda}). The same reasoning
can be applied to the dual map ${\tilde \Lambda}= I \otimes \Gamma$
and to the symmetric map $M=\Gamma\otimes\Gamma$. 
Thus, even if the maps $\Lambda'$ and $\Lambda''$
reveal inseparability by themselves, the combined map, 
Eq.~(\ref{eq_dilutionlambda}), is not guaranteed to do so because
the non-positivity of $(\Lambda' \otimes \Lambda'')\rho$ can be masked by
the two following terms.

\subsection{Special case of two 2-dimensional systems}

\noindent
{\bf Lemma 4:} 
The map $\Gamma$ acting on the state of a two-dimensional
system corresponds to time-reversal, and
is therefore equivalent to applying the complex conjugation operator $K$ 
followed by a rotation ${\cal R}_y$ by an angle $\pi$ about the
$y$-axis, that is, $\Gamma= {\cal R}_y K$.
\par
\medskip

Let us start by summarizing the action of the map
$\Gamma:\rho \to ({\rm Tr}\rho)-\rho$ on the density operator $\rho$
characterizing a 2-dimensional system (e.g., a quantum bit).
Since $\rho$ can be written
as a linear combination of the unit matrix and the three Pauli
matrices $\vec{\sigma}$ with {\em real} coefficients, 
it is sufficient to consider the action of $\Gamma$ on these (Hermitian)
basis matrices.
It is straightforward to check that $\Gamma$ is an antiunitary operator
that leaves the unit matrix unchanged
and flips the sign of the Pauli matrices $\sigma_{x,y,z}$,
\begin{equation}  \label{eq_mapGamma}
{\bf 1} \stackrel{\Gamma}{\longrightarrow} {\bf 1} \qquad
\sigma_x \stackrel{\Gamma}{\longrightarrow} -\sigma_x \qquad
\sigma_y \stackrel{\Gamma}{\longrightarrow} -\sigma_y \qquad
\sigma_z \stackrel{\Gamma}{\longrightarrow} -\sigma_z 
\end{equation}
so that, in the Bloch-sphere picture, we obtain
\begin{equation}  \label{eq_bloch2}
\rho={1\over 2} (1+\vec{r}\cdot \vec{\sigma}) \to
\Gamma \rho = 1 - \rho 
            = {1\over 2} (1-\vec{r}\cdot \vec{\sigma})
\end{equation}
where $\vec{r}$ is the Bloch vector.
Thus, the map $\Gamma$ flips $\vec{\sigma}$
(or performs a {\em parity} transformation on the
Bloch vector $\vec{r} \to - \vec{r}$), and corresponds
to {\em time-reversal}. It can therefore
be decomposed into complex conjugation $K$ followed by
a rotation ${\cal R}_y$ of an angle $\pi$ about the $y$-axis,
that is $\Gamma= T= {\cal R}_y K$~\cite{bib_schiff}. 
$\Box$
\par

\medskip
\noindent {\bf Remark:} 
The complex conjugation operator $K$ (or equivalently the
transposition, as we deal with Hermitian operators) 
corresponds to an {\em antiunitary} operator which acts on the four
basis matrices as:
\begin{equation}  \label{eq_mapK}
{\bf 1} \stackrel{K}{\longrightarrow} {\bf 1} \qquad
\sigma_x \stackrel{K}{\longrightarrow} \sigma_x \qquad
\sigma_y \stackrel{K}{\longrightarrow} -\sigma_y \qquad
\sigma_z \stackrel{K}{\longrightarrow} \sigma_z 
\end{equation}
(Remember that it is enough to consider the action of $K$ on the
basis matrices as the coefficients are real.)
Also, ${\cal R}_y$ is a unitary operation 
characterized by the unitary matrix
$U_y = \exp(-i\pi \sigma_y/2)=-i\sigma_y=\sigma_x\sigma_z$ 
which maps $\rho$ into $U_y \rho U_y^{\dagger}=\sigma_y\rho\sigma_y$,
so that the basis matrices are transformed according to
\begin{equation}   \label{eq_mapRy}
{\bf 1} \stackrel{{\cal R}_y}{\longrightarrow} {\bf 1} \qquad
\sigma_x \stackrel{{\cal R}_y}{\longrightarrow} -\sigma_x \qquad
\sigma_y \stackrel{{\cal R}_y}{\longrightarrow} \sigma_y \qquad
\sigma_z \stackrel{{\cal R}_y}{\longrightarrow} -\sigma_z 
\end{equation}
It is straightforward to check, using Eqs.~(\ref{eq_mapGamma}),
(\ref{eq_mapK}) and (\ref{eq_mapRy}),
that $\Gamma$ is the product of $K$ and ${\cal R}_y$.
(It is a general property of an antiunitary transformation that it
can be written as the product of a unitary transformation and a fixed
antiunitary operator such as time-reversal.) 
\par

In short, one can see that, if the system is in a (pure or mixed) state
given by Eq.~(\ref{eq_bloch}), then
\begin{equation}
U_y \rho^* U_y^{\dagger} = \sigma_y \rho^* \sigma_y
 = {1\over 2} (1+\sigma_y (\vec{r}\cdot \vec{\sigma}^*) \sigma_y)
 = {1\over 2} (1-\vec{r}\cdot \vec{\sigma})
 = \Gamma \rho
\end{equation}
where we have used the fact that $\vec{r}$ is a {\em real} vector and
that $\sigma_y \vec{\sigma} \sigma_y = - \vec{\sigma}^*$.
This generalizes what was shown earlier for pure states,
namely that if $|a\rangle= \alpha|0\rangle+\beta|1\rangle$ and
$|a^{\perp}\rangle = U_y(\alpha^*|0\rangle+\beta^*|1\rangle)=
 - \beta^*|0\rangle +\alpha^*|1\rangle$,
then we have
\begin{equation}
|a^{\perp}\rangle\langle a^{\perp}| = \Gamma (|a\rangle\langle a|)
\end{equation}

\noindent
{\bf Corollary 1:} When writing the Hilbert-Schmidt decomposition of
$\rho_{AB}$ for two 2-state systems,
the map $\Lambda=\Gamma\otimes I$ corresponds to a sign-flip
of the Pauli matrices acting on $A$ while leaving the sign of
those acting on $B$ unchanged.
\par
\medskip

Let us consider the Hilbert-Schmidt decomposition of an
arbitrary state of $AB$ (e.g., two quantum bits or
spin-1/2 particles)~\cite{bib_hor_alphaentropy}:
\begin{equation}  \label{eq_HilbertSchmidt}
\rho_{AB} = {1\over 4} \left( {\bf 1}_A \otimes {\bf 1}_B
+ \vec{r}\cdot \vec{\sigma}_A \otimes {\bf 1}_B
+ {\bf 1}_A \otimes \vec{s}\cdot \vec{\sigma}_B 
+ \sum_{m,n=1}^3 t_{n,m}\; \sigma_A^{(n)} \otimes \sigma_B^{(m)} \right)
\end{equation}
where $\sigma_A^{(n)}$ and $\sigma_B^{(m)}$
stand for the Pauli matrices (with $n=1,2,3$)
in the $A$ and $B$ space, respectively. Eq.~(\ref{eq_HilbertSchmidt})
depends on 15 real parameters, the two 3-dimensional vectors
$\vec{r}$ and $\vec{s}$, and the $3 \times 3$ real matrix $t_{n,m}$.
The vectors $\vec{r}$ and $\vec{s}$ correspond to the
state of $A$ and $B$ in the Bloch sphere since we have
\begin{eqnarray}
\rho_A &=& {\rm Tr}_B [\rho_{AB}]
 = {1\over 2} ({\bf 1}_A + \vec{r}\cdot \vec{\sigma}_A)  \\
\rho_B &=& {\rm Tr}_A [\rho_{AB}]
 = {1\over 2} ({\bf 1}_B + \vec{s}\cdot \vec{\sigma}_B)
\end{eqnarray}
They characterize the reduced systems $A$ and $B$, that is the
local (marginal) statistics of any observable on $A$ or $B$. The matrix 
$t_{n,m}={\rm Tr}[\rho_{AB} (\sigma_A^{(n)} \otimes \sigma_B^{(m)})]$
describes the joint statistics of $A$ and $B$, and yields for example
a very simple expression for the correlation between the measured
spin components along two axes (defined by the vectors 
$\vec{a}$ and $\vec{b}$):
\begin{equation}
{\rm Tr}\left[ \rho 
(\vec{a} \cdot \vec{\sigma}_A \otimes
 \vec{b} \cdot \vec{\sigma}_B ) \right]
= ( \vec{a}, t \vec{b})
\end{equation}
It is checked by straightforward calculation
that the map $\Lambda=\Gamma \otimes I$ 
simply flips the sign of the terms in $\vec{\sigma}_A$:
\begin{eqnarray}
\lambda_{AB} &=& {\bf 1}_A \otimes \rho_B - \rho_{AB}  \nonumber\\
&=& {1\over 4} \left( {\bf 1}_A \otimes {\bf 1}_B
- \vec{r}\cdot \vec{\sigma}_A \otimes {\bf 1}_B
+ {\bf 1}_A \otimes \vec{s}\cdot \vec{\sigma}_B 
- \sum_{m,n=1}^3 t_{n,m}\; \sigma_A^{(n)} \otimes \sigma_B^{(m)} \right)
\qquad \Box
\end{eqnarray}
\par

\medskip
\noindent
{\bf Corollary 2:} The dual map ${\tilde \Lambda}=I\otimes \Gamma$ 
flips the sign of the Pauli matrices acting on $B$ while leaving the sign of
those acting on $A$ unchanged. The action of the symmetric map $M$ on
the Hilbert-Schmidt decomposition of $\rho_{AB}$ is to
flip the sign of the Pauli matrices acting on both $A$ and $B$:
\begin{eqnarray}
\mu_{AB} &=& {\bf 1}_A \otimes {\bf 1}_B
- \rho_A \otimes {\bf 1}_B - {\bf 1}_A \otimes \rho_B + \rho_{AB}
\nonumber\\
&=& {1\over 4} \left( {\bf 1}_A \otimes {\bf 1}_B
- \vec{r}\cdot \vec{\sigma}_A \otimes {\bf 1}_B
- {\bf 1}_A \otimes \vec{s}\cdot \vec{\sigma}_B 
+ \sum_{m,n=1}^3 t_{n,m}\; \sigma_A^{(n)} \otimes \sigma_B^{(m)} \right)
\end{eqnarray}
This operation corresponds to time-reversal applied to $A$ and $B$
simultaneously, and can be shown to be equivalent to the 
conjugation operation in the ``magic'' basis
introduced by Hill and Wootters~\cite{bib_hillwootters} (see below).
\par

It is worth noting that the set of states that
remain invariant under the symmetric map $M$ are mixtures of
generalized Bell states, the latter being defined as the states obtained by
applying any local transformation to the four Bell states. These
states are also called ``$T$-states'' by 
Horodecki et al.~\cite{bib_hor_alphaentropy}, and
are such that the entropy of $A$ and $B$ is maximal, that is
$S(\rho_A)=S(\rho_B)=1$. (The only pure states in this set are
the fully entangled states of two qubits, i.e., the generalized Bell
states.) Thus, in particular, the (generalized) Bell states
are left unchanged by the action of $M$. In contrast,
a (separable) product state $\rho_A\otimes \rho_B$ is mapped into
the distinct state 
$\mu_{AB}=({\bf 1}_A - \rho_A)\otimes ({\bf 1}_B - \rho_B)$.
Because of this property, $\mu_{AB}$ by itself
is uninteresting as far as revealing inseparability
is concerned, as mentioned earlier.
\par

\medskip
\noindent
{\bf Theorem 7:} For two-dimensional systems $A$ and $B$,
the map $M$ conserves the spectrum, so that the separability
criteria resulting from the map $\Lambda$ and its dual ${\tilde \Lambda}$
are equivalent.
\par
\medskip

As we have seen before,
the map $\Gamma$ acting on subsystem $A$ (or $B$) amounts to
performing a complex conjugation operation $K$ followed by
a rotation ${\cal R}_y$ defined by $U_y=\exp(-i\pi\sigma_y/2)=-i\sigma_y$ 
(i.e., a bit- and phase-flip). Thus, the symmetric map $M=\Gamma\otimes\Gamma$
is equivalent to a complex conjugation $K$ (or transposition)
of the {\em joint} density operator in the Hilbert space ${\cal H}_{AB}$, 
followed by a tensor product of rotations 
$U_y\otimes U_y=-\sigma_y\otimes\sigma_y$.
Note that, as we are dealing with Hermitian (density) operators,
their spectrum is unchanged by $K$. The same is true for 
the rotation $U_y\otimes U_y$. As a consequence,
the operator $\mu_{AB}=M\rho_{AB}$
has the same spectrum as $\rho_{AB}$ when $d_A=d_B=2$. 
As $\Gamma$ is self-inverse ($\Gamma^2=I$) when $d_A=d_B=2$, 
we have the relation $I \otimes \Gamma = (\Gamma \otimes I) 
(\Gamma \otimes \Gamma)$ or in short ${\tilde \Lambda}=\Lambda M$.
This implies that
\begin{equation}
{\tilde \Lambda} \rho_{AB} = \Lambda \left[(U_y\otimes U_y)\rho_{AB}^*
(U_y^{\dagger}\otimes U_y^{\dagger})\right]
\end{equation}
which in turn results in
\begin{equation}
{\tilde \lambda}_{AB}= (U_y\otimes U_y) \lambda_{AB}^*
(U_y^{\dagger}\otimes U_y^{\dagger})
\end{equation}
as $\Lambda$ commutes with $U_y\otimes U_y$
and complex conjugation.
Since $\lambda_{AB}$ is Hermitian (just as $\rho_{AB}$), the latter
expression shows that the spectrum of ${\tilde \lambda}_{AB}$
and $\lambda_{AB}$ are identical,
so that the resulting criteria for separability are equivalent.
$\Box$
\par

\medskip
\noindent
{\bf Theorem 8:} For two-dimensional systems $A$ and $B$
which have maximal reduced entropy, i.e., $S(\rho_A)=S(\rho_B)=1$, the
positivity of the operator $\Lambda \rho_{AB}$ results 
in a necessary {\em and} sufficient condition for the
separability of $\rho_{AB}$.
\par
\medskip

The states with $\vec{r}=\vec{s}=0$ are such that the reduced
density operators are given by $\rho_A=\rho_B={\bf 1}/2$, so that
the reduced entropies are $S(\rho_A)=S(\rho_B)=1$. These ``$T$-states'' 
are thus completely characterized by the matrix $t_{n,m}$. It has been
shown in Ref.~\cite{bib_hor_alphaentropy} that any state belonging 
to this set of $T$-states can be transformed by a unitary transformation
of the product form $U_A \otimes U_B$ into a state for which
$t_{n,m}$ is {\em diagonal}. As far as separability
is concerned, we can thus restrict ourselves to the class of
all states with diagonal $t$, since these are representative
of the entire set of $T$-states (up to an $U_A \otimes U_B$ isomorphism).
\par

The class of states with diagonal $t$ is a convex
subset of the set of $T$-states, and any state belonging to this
subset can be characterized by the real vector 
$\vec{t}=(t_{11},t_{22},t_{33})$
made out of the diagonal elements of $t$.
It has been shown in Ref.~\cite{bib_hor_alphaentropy}
that an operator $\rho_{AB}$ of the form
given by Eq.~(\ref{eq_HilbertSchmidt}) with $\vec{r}=\vec{s}=0$
and diagonal $t$ corresponds to a state (i.e., a {\em positive}
unit-trace operator) if and only if the vector $\vec{t}$ belongs
to a tetrahedron with vertices $\vec{t}_1=(-1,1,1)$,
$\vec{t}_2=(1,-1,1)$, $\vec{t}_3=(1,1,-1)$, and $\vec{t}_4=(-1,-1,-1)$.
In other words, any state of this class 
can be represented by a point inside this tetrahedron.
In this representation, the four Bell states
$|\Phi^{\pm}\rangle = 2^{-1/2} (|00\rangle \pm |11\rangle)$ and
$|\Psi^{\pm}\rangle = 2^{-1/2} (|01\rangle \pm |10\rangle)$
correspond to the vertices of the tetrahedron, that is
\begin{eqnarray}
\vec{t}_1 &:& \qquad |\Phi^-\rangle \langle \Phi^-| =  
{1\over 4} \left( {\bf 1}_A \otimes {\bf 1}_B
- \sigma_A^{(x)} \otimes \sigma_B^{(x)}
+ \sigma_A^{(y)} \otimes \sigma_B^{(y)}
+ \sigma_A^{(z)} \otimes \sigma_B^{(z)} \right)
\nonumber \\
\vec{t}_2 &:& \qquad |\Phi^+\rangle \langle \Phi^+| =  
{1\over 4} \left( {\bf 1}_A \otimes {\bf 1}_B
+ \sigma_A^{(x)} \otimes \sigma_B^{(x)}
- \sigma_A^{(y)} \otimes \sigma_B^{(y)}
+ \sigma_A^{(z)} \otimes \sigma_B^{(z)} \right)
\nonumber \\
\vec{t}_3 &:& \qquad |\Psi^+\rangle \langle \Psi^+| =  
{1\over 4} \left( {\bf 1}_A \otimes {\bf 1}_B
+ \sigma_A^{(x)} \otimes \sigma_B^{(x)}
+ \sigma_A^{(y)} \otimes \sigma_B^{(y)}
- \sigma_A^{(z)} \otimes \sigma_B^{(z)} \right)
\nonumber \\
\vec{t}_4 &:& \qquad |\Psi^-\rangle \langle \Psi^-| =  
{1\over 4} \left( {\bf 1}_A \otimes {\bf 1}_B
- \sigma_A^{(x)} \otimes \sigma_B^{(x)}
- \sigma_A^{(y)} \otimes \sigma_B^{(y)}
- \sigma_A^{(z)} \otimes \sigma_B^{(z)} \right)
\end{eqnarray}
In Ref.~\cite{bib_hor_alphaentropy}, it is also shown that a state $\rho_{AB}$
of this $T$-diagonal class is {\em separable} if and only if 
the vector $\vec{t}$ characterizing $\rho_{AB}$ belongs to an
octahedron with vertices $\vec{o}_1^{\;\pm}=(\pm 1,0,0)$,
$\vec{o}_2^{\;\pm}=(0,\pm 1,0)$, and
$\vec{o}_3^{\;\pm}=(0,0,\pm 1)$. This results in
a necessary {\em and} sufficient condition for separability 
within the class of $T$-states.
\par

Let us consider the action of $\Lambda$ in this
representation. As shown earlier, $\Lambda$ flips the ``spin''
$\vec{\sigma}_A$. Within the set of $T$-states,
this amounts to changing the sign of the $t_{n,m}$ matrix, that
is, to flipping the sign of the vector $\vec{t}$
for $T$-diagonal states. Therefore, the criterion for
separability $\lambda_{AB}=\Lambda \rho_{AB} \ge 0$
translates, in this representation, to the condition that the
``parity'' operation on the vector $\vec{t}$ characterizing a 
separable state results in a positive operator (i.e., a legitimate state).
Thus, $-\vec{t}$ must belong to the tetrahedron. It is easy to see
that the set of points of the tetrahedron which are such that their
image under parity still belongs to the tetrahedron
corresponds exactly to the octahedron defined above. Therefore,
no inseparable state exists that satisfies $\Lambda \rho_{AB} \ge 0$,
so that $\Lambda$ provides a necessary {\em and}  sufficient
condition for separability within the class of $T$-states. 
$\Box$
\par

\medskip
\noindent
{\bf Theorem 9:} A bipartite system of two-dimensional components $A$ and $B$
characterized by an arbitrary joint density operator $\rho_{AB}$ is
separable {\em if and only if} the operator 
$\lambda_{AB}= \Lambda \rho_{AB}$ is positive semi-definite.
\par
\medskip

It is enough to show that
$\Lambda$ is equivalent to a partial transposition
up to some completely positive map (in fact, a unitary transformation). 
As already mentioned,
Peres' separability criterion is based on the partial transposition
operation, that is, on the map $T \otimes I$, where $T$ is the
standard transposition of operators on the ${\cal H}_A$ subspace.
Since we are dealing with Hermitian operators, 
$T \otimes I$ is equivalent to the ``partial conjugation'' operation
$K \otimes I$, where $K$ is the complex conjugation operator 
acting on states on ${\cal H}_A$. Note that, although $K$ is
well-defined, partial conjugation $K \otimes I$
is {\em only} defined for product state vectors
in ${\cal H}_{AB}$~\cite{bib_horo_one}.
We can now use Lemma 4, i.e., $\Gamma = {\cal R}_y K$,
together with the fact that any positive map $\Pi$ 
acting on density operators in a two-dimensional
Hilbert space can be written as~\cite{bib_horo_sufficient} 
\begin{equation}
\Pi = \Pi_1^{\rm CP} + \Pi_2^{\rm CP} T
\end{equation}
where $\Pi_1^{\rm CP}$ and $\Pi_2^{\rm CP}$ are completely
positive maps (which therefore do not reveal inseparability).
With the identification $\Pi_1^{\rm CP}=0$ and 
$\Pi_2^{\rm CP}={\cal R}_y$, we see that the map $\Gamma$ can be used
rather than the transposition operator $T$ (or $K$) in order to test
the positivity of the operator resulting from applying 
{\em any} element of the set of maps $\Pi \otimes I$ on $\rho_{AB}$.
This is simply due to the fact that the complex conjugation operator $K$
is {\em unitarily} equivalent to $\Gamma$.
Thus, the reasoning used in Ref.~\cite{bib_horo_sufficient}
holds here, so that $\Lambda \rho_{AB}\ge 0$ results 
in a necessary {\em and} sufficient condition for the
separability of $\rho_{AB}$.
$\Box$
\par

\medskip
\noindent {\bf Remark 1:}
Since the spectrum of an operator is
conserved by a unitary transformation (${\cal R}_y$), it
is clear that the spectrum of the matrix obtained
by partial transposition in subspace $A$, $\rho_{AB}^{T_A}$,
is the same as the spectrum of $\lambda_{AB}=\Lambda\rho_{AB}$.
Therefore, testing Peres' condition or the positivity of
$\lambda_{AB}$ is operationally equivalent, and these conditions can be
used interchangeably in the case of two quantum bits, as illustrated in
Appendix A. Since $\Gamma$ corresponds to
time-reversal (on the subsystem $A$),
the map $\Lambda=\Gamma\otimes I$ amounts to changing the arrow
of time for subsystem $A$ with respect to subsystem $B$. Such a relation
between time-reversal and Peres' map has been pointed out previously
by Sanpera et al.~\cite{bib_sanpera}, where it was shown that the
partial transposition operator is unitarily equivalent to 
``local'' time-reversal.
\par

\medskip
\noindent {\bf Remark 2:}
As we know that $\Gamma$ applied to a two-dimensional
system is unitarily equivalent to the transposition operator $T$,
the separability condition based on $\Lambda=\Gamma\otimes I$
is in fact equivalent to the one based on the partial transpose 
$(T\otimes I)\rho_{AB}=\rho_{AB}^{T_A}$
whenever subsystem $A$ is two-dimensional. More precisely,
$\lambda_{AB}$ and $\rho_{AB}^{T_A}$ have the same spectrum
for $2\times n$ systems, so that the conditions
are equivalent if $\Gamma$ is applied on the two-dimensional
subsystem. As a consequence, the separability condition
based on $\Lambda$ is also necessary {\em and} sufficient for
$2\times 3$ systems, while it is only necessary for 
$2\times n$ systems with larger $n$, just as Peres'
condition~\cite{bib_horo_sufficient}.
Numerical evidence suggests that, for systems 
with $d_A,d_B>2$, the condition based on $\Lambda$ 
(or ${\tilde\Lambda}$) is weaker than (or equivalent to)
the one based on partial transposition.

\medskip
\noindent
{\bf Theorem 10:} The symmetric map $M=\Gamma\otimes\Gamma$ applied
to a bipartite system of two-dimensional components (i.e., global
time-reversal) is equivalent to complex conjugation
in the ``magic'' basis introduced in Ref.~\cite{bib_bdsw}.
\par
\medskip

Since $\Gamma={\cal R}_y K$, where $K$ denotes
the conjugation operator and ${\cal R}_y$ is a rotation 
characterized by $U_y=\exp(-i\pi\sigma_y/2)=-i\sigma_y$,
the symmetric map $M$ applied to the state $\rho_{AB}$
of a bipartite systems results in
\begin{equation}  \label{eq_magic1}
M \rho_{AB} = (U_y \otimes U_y) \rho_{AB}^* 
(U_y^{\dagger} \otimes U_y^{\dagger})
\end{equation}
where $U_y \otimes U_y=-\sigma_y\otimes \sigma_y$.
Since $M$ is antiunitary and self-inverse ($M^2=I$), it is
a {\em conjugation}~\cite{bib_reedsimon}. It can be written
as the complex conjugation operator if expressed in a specific basis. 
Let us assume that $V$ is the unitary operator (in the joint space) 
that transforms the product states into the states $\{|e_i\rangle\}$
that form this specific basis, that is
\begin{equation}
|e_1\rangle= V|00\rangle \qquad
|e_2\rangle= V|01\rangle \qquad
|e_3\rangle= V|10\rangle \qquad
|e_4\rangle= V|11\rangle
\end{equation}
We would like to show that $M$ is equivalent to rotating the
states $|e_i\rangle$ into the product states, taking the complex
conjugation of the density matrix (in the product basis), 
and then rotating the product states back to the $|e_i\rangle$'s:
\begin{equation}  \label{eq_magic2}
M \rho_{AB} = V (V^{\dagger} \rho_{AB} V)^* V^{\dagger}
            = (V V^T) \rho_{AB}^* (V V^T)^{\dagger}
\end{equation}
where $V^T$ is the transpose of the unitary matrix $V$.
Identifying Eqs.~(\ref{eq_magic1}) and (\ref{eq_magic2}), we obtain
\begin{equation}  \label{eq_solveV}
V V^T = U_y \otimes U_y = - \sigma_y \otimes \sigma_y
= \left( \begin{array}{cccc}
 0 & 0 & 0 & 1 \\
 0 & 0 &-1 & 0 \\
 0 &-1 & 0 & 0 \\
 1 & 0 & 0 & 0
\end{array} \right)
\end{equation}
It is easy to prove that, if $V$ is unitary, then $V V^T$ is
unitary and symmetric (but not necessarily Hermitian).
In order to find a solution for $V$ that satisfies
Eq.~(\ref{eq_solveV}), we first diagonalize the matrix 
$\sigma_y\otimes\sigma_y$. Consider the unitary matrix
\begin{equation}
W \equiv \exp\left(-{i\pi\over 4} (1-\sigma_x)\otimes(1-\sigma_x)\right)
 = (1 \otimes 1 + 1 \otimes \sigma_x + \sigma_x \otimes 1 
    - \sigma_x \otimes \sigma_x)/2
\end{equation}
This matrix is self-inverse, that is, $W^2=1$, so that
it is Hermitian ($W^{\dagger} = W$). It can easily be shown that
it diagonalizes $\sigma_y\otimes\sigma_y$, i.e.,
\begin{equation}
W (\sigma_y\otimes\sigma_y) W = \sigma_z \otimes \sigma_z
\end{equation}
Replacing $V$ in Eq.~(\ref{eq_solveV}) by the product of a diagonal
matrix $D$ and $W$, that is $V=WD$, we obtain
\begin{equation}
DD^T=-W(\sigma_y\otimes\sigma_y)W=-\sigma_z \otimes \sigma_z
= \left( \begin{array}{cccc}
 -1 & 0 & 0 & 0 \\
 0 & 1 & 0 & 0 \\
 0 & 0 & 1 & 0 \\
 0 & 0 & 0 &-1
\end{array} \right)
\end{equation}
which implies that
\begin{equation}
D=\left( \begin{array}{cccc}
 \pm i & 0 & 0 & 0 \\
 0 & \pm 1 & 0 & 0 \\
 0 & 0 & \pm 1 & 0 \\
 0 & 0 & 0 & \pm i
\end{array} \right)
\end{equation}
This yields a (non-unique) solution for the unitary matrix 
$V=WD$ which defines
the basis $\{ |e_i\rangle \}$. It is worth noticing that the matrix
\begin{equation}
W= {1 \over 2} \left( \begin{array}{cccc}
 1 & 1 & 1 & -1 \\
 1 & 1 & -1 & 1 \\
 1 & -1 & 1 & 1 \\
 -1 & 1 & 1 & 1
\end{array} \right)
\end{equation}
transforms the product states
into the four maximally entangled states which are obtained by
applying a local transformation $H\otimes 1$ on the four Bell
states, i.e.,
\begin{eqnarray}
W|00\rangle &=& (|00\rangle+|01\rangle+|10\rangle-|11\rangle)/2 
= (H\otimes 1) |\Phi^+\rangle \nonumber\\
W|01\rangle &=& (|00\rangle+|01\rangle-|10\rangle+|11\rangle)/2 
= (H\otimes 1) |\Psi^+\rangle  \nonumber\\
W|10\rangle &=& (|00\rangle-|01\rangle+|10\rangle+|11\rangle)/2 
= (H\otimes 1) |\Phi^-\rangle  \nonumber\\
W|11\rangle &=& (-|00\rangle+|01\rangle+|10\rangle+|11\rangle)/2 
= (H\otimes 1) |\Psi^-\rangle
\end{eqnarray}
where $H$ is the Hadamard transform. (As a matter of fact, the unitary
transformation $W$ corresponds simply to a controlled-{\sc not} gate
where the control is in the dual basis $\{|0\rangle+|1\rangle,
|0\rangle-|1\rangle\}$ rather than the standard basis.)
Therefore, the unitary transformation
$V=WD$ is such that the product states are rotated into the four
generalized Bell states with the appropriate phases
\begin{eqnarray}
|e_1\rangle = V|00\rangle &=& \pm i (H\otimes 1) |\Phi^+\rangle \nonumber\\
|e_2\rangle = V|01\rangle &=& \pm 1 (H\otimes 1) |\Psi^+\rangle \nonumber\\
|e_3\rangle = V|10\rangle &=& \pm 1 (H\otimes 1) |\Phi^-\rangle \nonumber\\
|e_4\rangle = V|11\rangle &=& \pm i (H\otimes 1) |\Psi^-\rangle \nonumber\\
\end{eqnarray}
These states $|e_i\rangle$ are therefore equivalent, up to a
local change of basis $H\otimes 1$ and a phase $i$ that are irrelevant here, 
to the ``magic'' states introduced in
Ref.~\cite{bib_bdsw}. (Any four states
obtained from the $|e_i\rangle$'s up to an overall phase and
a unitary transformation acting locally on each quantum bit
are legitimate ``magic'' states.) This implies that, when expressed in
this basis, the symmetric map $M=\Gamma\otimes\Gamma$ 
reduces the the complex conjugation operation that was used
in the context of the calculation of 
the entropy of formation of a pair of quantum bits
(see Refs.~\cite{bib_hillwootters,bib_wootters}).
$\Box$
\par

\section{Conclusion}

Given a bipartite system characterized
by a density operator $\rho_{AB}$, one can define a
conditional amplitude operator $\rho_{A|B}$ (a positive Hermitian
operator defined on the range of $\rho_{AB}$) which
plays the role of a conditional probability in quantum information
theory. Specifically, it can be used to define a conditional von Neumann
entropy, $S(A|B)=-{\rm Tr}[\rho_{AB} \log \rho_{A|B}]$, in perfect
analogy with the conditional Shannon entropy. 
The quantum counterpart of many classical properties
also holds: i)~$\rho_{A|B}$ is defined on the support of $\rho_{AB}$,
so that $S(A|B)$ is well-defined; ii)~$S(A|B)=S(AB)-S(B)$; 
iii)~$\rho_{A|B}=\rho_A \otimes {\bf 1}_B$ if $A$ and $B$ are
independent; 
iv)~$\rho_{AA'|BB'}=\rho_{A|B}\otimes \rho_{A'|B'}$
if $\rho_{AA';BB'}=\rho_{AB}\otimes\rho_{A'B'}$;
v)~$\rho_{A|B}$ transforms as 
$(U_A\otimes U_B)\rho_{A|B}(U_A^{\dagger}\otimes U_B^{\dagger})$
when performing a local unitary transformation $U_A\otimes U_B$ on
$\rho_{AB}$, so that its spectrum and therefore $S(A|B)$ are invariant
under such transformations on $\rho_{AB}$.
\par

The main non-classical feature that appears when dealing
with a quantum bipartite system rather than a classical one
is that $\rho_{A|B}$ may have a ``non-classical'' spectrum,
that is, eigenvalues of $\rho_{A|B}$ may exceed 1, which in turn
implies that $S(A|B)$ can be negative. More specifically, we have shown
that $\rho_{AB} \le 1$ for any separable state, which also
straightforwardly implies $S(A|B) \ge 0$. Therefore, a {\em necessary}
condition for separability is that the conditional amplitude operator
has a ``classical'' spectrum, or that the conditional
entropy is non-negative (the latter is a weaker condition). These
conditions are not sufficient, since extending an inseparable state with
a separable one of large dimension may result in a {\em dilution}
of inseparability, that is, it may give rise to a state with
$\rho_{A|B} \le 1$. In other words, some
inseparable states exist with $\rho_{A|B} \le 1$, and certainly some
with $S(A|B)\ge 0$ (even if $\rho_{A|B}\not\le 1$).
\par

These considerations can be used to define a simpler necessary
condition for separability, based on the positive linear
map $\Gamma:\rho \to ({\rm Tr}\rho) -\rho$. Any separable state
is mapped by the tensor product of $\Gamma$ (acting on $A$)
and the identity $I$ (acting on $B$) into a positive operator.
Therefore, a simple separability criterion is based on checking
the positivity of the operator 
$(\Gamma \otimes I)\rho_{AB} = {\bf 1}_A \otimes \rho_B - \rho_{AB}$.
This condition, along with the one based on the dual
map $I\otimes \Gamma$, can be shown to be non-sufficient
for a system of arbitrary dimension. 
Since the map $\Gamma$ commutes with any unitary transformation, the
spectrum of the operator $(\Gamma \otimes I)\rho_{AB}$ is invariant
under a local unitary transformation $U_A\otimes U_B$, making this
condition independent of the basis in which $A$ and $B$ are expressed.
\par

In the case of a 2-dimensional system, the map $\Gamma$ can be shown
to be the time-reversal operator, which flips the sign of the
spin matrices (or, consequently, reverses the Bloch vector 
characterizing the state of the quantum bit). It is therefore unitarily
equivalent to the transposition operator~\cite{bib_peres,bib_sanpera},
so that the
condition based on $\Gamma\otimes I$ is equivalent to the one based
on Peres' partial transposition
for $2 \times n$ systems (when applying the map $\Gamma$ on the
2-dimensional subsystem). As a
consequence, it is necessary {\em and} sufficient for
$2\times 2$ and $2\times 3$ systems while it is only necessary
for larger systems, just as is Peres'~\cite{bib_horo_sufficient}. 
Numerical evidence suggests that, for systems 
where $d_A,d_B>2$, the condition based
on $\Gamma\otimes I$ or $I\otimes \Gamma$ is weaker than (or equivalent to)
the one based on partial transposition.
\par

Finally, we consider the symmetric map, defined as
$(\Gamma\otimes\Gamma)\rho_{AB}= {\bf 1}_A \otimes {\bf 1}_B
-\rho_A \otimes {\bf 1}_B - {\bf 1}_A \otimes \rho_B + \rho_{AB}$.
The states which are left invariant under this map
are mixtures of generalized Bell states (or ``T-states''),
which include the maximally entangled pure states as well as the product
of two independent (unentangled) random bits. 
It can be seen that the map $\Gamma\otimes\Gamma$ 
is also related to quantum nonlocality of two quantum bits
even though it does not directly reveal inseparability.
Indeed, $\Gamma\otimes\Gamma$ 
reduces to the complex conjugation in the ``magic'' basis
that has been used in the context of the calculation of 
the entropy of formation of a pair of quantum bits
(see Refs.~\cite{bib_hillwootters,bib_wootters}). It therefore might be
interesting to look for a simple relation between the map $\Gamma$ (related
to inseparability) and the entropy of formation.
This will be the subject of further work.

\bigskip

\noindent {\it Note:} Parts of Section IIIB of this paper are
equivalent to results contained in
Ref.~\cite{bib_wootters}, which was brought to our attention
after completion of this work.

\acknowledgments
We acknowledge useful discussions with Lev Levitin,
Barry Simon, and Armin Uhlmann.
We are also grateful to Chris Fuchs for communicating
to us unpublished results of Ref.~\cite{bib_wootters}, especially
the connection between the map $M$ and the ``magic'' basis for two qubits.
This work was supported in part by NSF Grants
PHY 94-12818 and PHY 94-20470, and by a grant from DARPA/ARO
through the QUIC Program (\#DAAH04-96-1-3086).



\appendix
\section{Examples}

Here we consider several examples illustrating the separability criterion
$\lambda_{AB}\ge 0$, and compare it to
Peres' criterion~\cite{bib_peres}. Examples 1-4 deal
with states of two quantum bits, and illustrate the fact
that the $\Lambda$-criterion is necessary and sufficient (the spectrum 
of $\lambda_{AB}$ is identical to the spectrum of $\rho^{T_A}$).
Examples 5-6 illustrate that
the $\Lambda$-condition is not sufficient for systems in
larger dimensions ($3\times 3$ and $2\times 4$) whose partial
transpose is positive (cf. Ref.~\cite{bib_horo_sufficient}). 
In fact,
the $\Lambda$-condition is equivalent to Peres' condition for $2\times n$
systems, so that it is also necessary and sufficient for $2\times 3$ 
systems~\cite{bib_horo_sufficient} while it is only necessary
for larger $n$.
\par

\medskip
\noindent {\bf Example 1}: Consider a Werner state~\cite{bib_werner}
with parameter $x$
($0\le x \le 1$), that is a mixture of a fraction $x$ of the singlet 
state $|\Psi^-\rangle$ and a random fraction $(1-x)$. We shall
see that $\lambda_{AB}\ge 0$ is equivalent to
Peres' criterion, and is therefore sufficient in this case.
Indeed, the joint density operator 
\begin{equation}
\rho_{AB}= x |\Psi^-\rangle \langle\Psi^-| 
+ {(1-x)\over 4} ({\bf 1} \otimes {\bf 1})
= \left( \begin{array}{cccc}
 {1-x\over 4} & 0 & 0 & 0 \\
 0 & {1+x\over 4} & -{x\over 2} & 0 \\
 0 & -{x\over 2} & {1+x\over 4} & 0 \\
 0 & 0 & 0 & {1-x\over 4}
\end{array} \right)
\end{equation}
is mapped by $\Lambda$ into the operator
\begin{equation}
\lambda_{AB}= \left( \begin{array}{cccc}
 {1+x\over 4} & 0 & 0 & 0 \\
 0 & {1-x\over 4} & {x\over 2} & 0 \\
 0 & {x\over 2} & {1-x\over 4} & 0 \\
 0 & 0 & 0 & {1+x\over 4}
\end{array} \right)
\end{equation}
which admits three eigenvalues equal to $(1+x)/4$ and a
fourth equal to $(1-3x)/4$. The latter becomes negative
if $x>1/3$, so that $\lambda_{AB}$ is
positive semi-definite only if $x\le 1/3$, which has been
proven to be the {\em exact} threshold for separability (any Werner state
with $x\le 1/3$ is separable as it can be written as a mixture 
of product states~\cite{bib_bbpssw}). As expected,
the spectrum of $\lambda_{AB}$ is equal
to the spectrum of the partial transpose of $\rho_{AB}$, so that
the $\Lambda$-condition is sufficient to ensure
separability for Werner states.
\par

\medskip
\noindent {\bf Example 2}: Consider a mixed state that is made
out of a fraction $x$ of the entangled state 
$|\psi\rangle=a|01\rangle+b|10\rangle$,
and fractions $(1-x)/2$ of the separable product states 
$|00\rangle$ and $|11\rangle$
(see~\cite{bib_gisin}). The joint density matrix is of the form
\begin{equation}
\rho_{AB}= x |\psi\rangle \langle\psi| 
+ {1-x\over 2} |00\rangle \langle 00|
+ {1-x\over 2} |11\rangle \langle 11|
= \left( \begin{array}{cccc}
 {1-x\over 2} & 0 & 0 & 0 \\
 0 & x|a|^2 & xab^* & 0 \\
 0 & xa^* b & x|b|^2 & 0 \\
 0 & 0 & 0 & {1-x\over 2}
\end{array} \right)
\end{equation}
with $a$ and $b$ satisfying $|a|^2+|b|^2=1$.
It is mapped by $\Lambda$ into the matrix
\begin{equation}
\lambda_{AB}= \left( \begin{array}{cccc}
x|b|^2   & 0 & 0 & 0 \\
 0 & {1-x\over 2} & -xab^* & 0 \\
 0 & -xa^* b & {1-x\over 2} & 0 \\
 0 & 0 & 0 & x|a|^2
\end{array} \right)
\end{equation}
The eigenvalues of $\lambda_{AB}$ are $x|a|^2$, $x|b|^2$, and
$(1-x\pm 2x|ab|)/2$. This implies that $\rho_{AB}$ is
inseparable if $x>(1+2|ab|)^{-1}$, exactly as predicted
by Peres using the partial transpose of $\rho_{AB}$. Since
we are dealing with two qubits, this is the exact limit
between separability and inseparability~\cite{bib_peres,bib_horo_sufficient}.
\par

\medskip
\noindent {\bf Example 3}: In the simpler case where $\rho_{AB}$ is
a mixture of a fraction $x$ of the singlet state $|\Psi^-\rangle$
and a fraction $(1-x)$ of the separable product state $|00\rangle$,
\begin{equation}
\rho_{AB}= x |\psi\rangle \langle\psi| 
+ (1-x) |00\rangle \langle 00|
= \left( \begin{array}{cccc}
 1-x & 0 & 0 & 0 \\
 0 & x/2 & -x/2 & 0 \\
 0 & -x/2 & x/2 & 0 \\
 0 & 0 & 0 & 0
\end{array} \right)
\end{equation}
we obtain
\begin{equation}
\lambda_{AB}= \left( \begin{array}{cccc}
x/2& 0 & 0 & 0 \\
 0 & 0 & x/2 & 0 \\
 0 & x/2 & 1-x & 0 \\
 0 & 0 & 0 & x/2
\end{array} \right)
\end{equation}
The latter matrix admits two eigenvalues equal to $x/2$ and
two eigenvalues equal to $\left(1-x\pm\sqrt{(1-x)^2+x^2}\right)\Big/2$, 
so that its determinant is equal to $-(x/2)^4$.
Thus, this state is inseparable whenever $x>0$, as expected. (It is
separable only if it is the pure product state $|00\rangle$.)
\par

\medskip
\noindent {\bf Example 4}: Consider the class of 
2-qubit inseparable states described by 
Horodecki et al.~\cite{bib_horo_sufficient},
a mixture of two entangled states:
\begin{equation}
\rho_{AB}= p |\psi_1\rangle \langle \psi_1| +
(1-p) |\psi_2\rangle \langle \psi_2|
\end{equation}
where $|\psi_1\rangle=a|00\rangle+b|11\rangle$ and
$|\psi_2\rangle=a|01\rangle+b|10\rangle$, with $a,b>0$ and 
satisfying $|a|^2+|b|^2=1$. The joint density matrix
\begin{equation}
\rho_{AB} = \left( \begin{array}{cccc}
 pa^2 & 0 & 0 & pab \\
 0 & (1-p)a^2 & (1-p)ab & 0 \\
 0 & (1-p)ab & (1-p)b^2 & 0 \\
 pab & 0 & 0 & pb^2
\end{array} \right)
\end{equation}
is mapped by $\Lambda$ to
\begin{equation}
\lambda_{AB} = \left( \begin{array}{cccc}
 (1-p)b^2 & 0 & 0 & -pab \\
 0 & pb^2 & (p-1)ab & 0 \\
 0 & (p-1)ab & pa^2 & 0 \\
 -pab & 0 & 0 & (1-p)a^2
\end{array} \right)
\end{equation}
The latter matrix admits two eigenvalues equal to
$\left(p\pm \sqrt{p^2+4 a^2 b^2 (1-2p)}\right)\Big/2$ and
two eigenvalues equal to
$\left(1-p\pm \sqrt{(1-p)^2+4 a^2 b^2 (2p-1)}\right)\Big/2$,
so that its determinant is equal to $-a^4 b^4 (1-2p)^2$.
This state is therefore inseparable whenever $ab\ne 0$
and $p\ne 1/2$, in perfect agreement with Ref.~\cite{bib_horo_sufficient}.
\par

\medskip
\noindent {\bf Example 5}: Consider the $3\times 3$ system
in a weakly inseparable state introduced 
by Horodecki~\cite{bib_horo_one},
\begin{equation}
\rho_{AB}=  {1 \over 1 + 8a}
\left( \begin{array}{ccccccccc}
 a & 0 & 0 & 0 & a & 0 & 0 & 0 & a \\
 0 & a & 0 & 0 & 0 & 0 & 0 & 0 & 0 \\
 0 & 0 & a & 0 & 0 & 0 & 0 & 0 & 0 \\
 0 & 0 & 0 & a & 0 & 0 & 0 & 0 & 0 \\
 a & 0 & 0 & 0 & a & 0 & 0 & 0 & a \\
 0 & 0 & 0 & 0 & 0 & a & 0 & 0 & 0 \\
 0 & 0 & 0 & 0 & 0 & 0 & {1+a\over 2} & 0 & {\sqrt{1-a^2}\over 2} \\
 0 & 0 & 0 & 0 & 0 & 0 & 0 & a & 0 \\
 a & 0 & 0 & 0 & a & 0 & {\sqrt{1-a^2}\over 2} & 0 & {1+a\over 2}
\end{array} \right)
\end{equation}
where $a$ is a parameter ($a\ne 0,1$). As shown in Ref.~\cite{bib_horo_one},
the partial transpose of this state is positive, although $\rho_{AB}$
is inseparable, which makes the inseparability of $\rho_{AB}$ undetectable
using Peres' criterion. It is simple to check that the 
$\Lambda$-mapped operator
\begin{equation}
\lambda_{AB}=  {1 \over 1 + 8a}
\left( \begin{array}{ccccccccc}
 {1+3a\over 2} & 0 & {\sqrt{1-a^2}\over 2} & 0 & -a & 0 & 0 & 0 & -a \\
 0 & 2a & 0 & 0 & 0 & 0 & 0 & 0 & 0 \\
 {\sqrt{1-a^2}\over 2} & 0 & {1+3a\over 2} & 0 & 0 & 0 & 0 & 0 & 0 \\
 0 & 0 & 0 & {1+3a\over 2} & 0 & {\sqrt{1-a^2}\over 2} & 0 & 0 & 0 \\
 -a & 0 & 0 & 0 & 2a & 0 & 0 & 0 & -a \\
 0 & 0 & 0 & {\sqrt{1-a^2}\over 2} & 0 & {1+3a\over 2} & 0 & 0 & 0 \\
 0 & 0 & 0 & 0 & 0 & 0 & 2a & 0 & 0 \\
 0 & 0 & 0 & 0 & 0 & 0 & 0 & 2a & 0 \\
 -a & 0 & 0 & 0 & -a & 0 & 0 & 0 & 2a
\end{array} \right)
\end{equation}
is positive (with a trace equal to 2),
so that $\Lambda$ cannot reveal the inseparability 
of $\rho_{AB}$ either.
Accordingly, the determinant of $\lambda_{AB}$ is equal to 
$6a^7(1-a)(5a+3)/(1+8a)^9$ and thus positive. Note that the dual
map also yields a positive operator ${\tilde \lambda}_{AB}$ (of trace 2), 
although the eigenvalues of ${\tilde \lambda}_{AB}$ are
distinct from those of $\lambda_{AB}$, as is its determinant
${\rm Det}({\tilde \lambda}_{AB})= 24a^7(1-a^2)/(1+8a)^9$.
This example emphasizes the fact that $\Lambda$ does not
result in a sufficient separability condition for $3\times 3$ systems,
just as Peres' condition~\cite{bib_horo_sufficient}.
\par

\medskip
\noindent {\bf Example 6}: Following Horodecki~\cite{bib_horo_one},
we consider a $2\times 4$ system in an inseparable state
\begin{equation}
\rho_{AB}=  {1 \over 1 + 7b}
\left( \begin{array}{cccccccc}
 b & 0 & 0 & 0 & 0 & b & 0 & 0 \\
 0 & b & 0 & 0 & 0 & 0 & b & 0 \\
 0 & 0 & b & 0 & 0 & 0 & 0 & b \\
 0 & 0 & 0 & b & 0 & 0 & 0 & 0 \\
 0 & 0 & 0 & 0 & {1+b\over 2} & 0 & 0 & {\sqrt{1-b^2}\over 2} \\
 b & 0 & 0 & 0 & 0 & b & 0 & 0 \\
 0 & b & 0 & 0 & 0 & 0 & b & 0 \\
 0 & 0 & b & 0 & {\sqrt{1-b^2}\over 2} & 0 & 0 & {1+b\over 2}
\end{array} \right)
\end{equation}
that has a positive partial transpose,
where $b$ is a parameter ($b\ne 0,1$). Applying $\Lambda$, we
see that
\begin{equation}
\lambda_{AB}=  {1 \over 1 + 7b}
\left( \begin{array}{cccccccc}
 {1+b\over 2} & 0 & 0 & {\sqrt{1-b^2}\over 2} & 0 & -b & 0 & 0 \\
 0 & b & 0 & 0 & 0 & 0 & -b & 0 \\
 0 & 0 & b & 0 & 0 & 0 & 0 & -b \\
 {\sqrt{1-b^2}\over 2} & 0 & 0 & {1+b\over 2} & 0 & 0 & 0 & 0 \\
 0 & 0 & 0 & 0 & b & 0 & 0 & 0 \\
 -b & 0 & 0 & 0 & 0 & b & 0 & 0 \\
 0 & -b & 0 & 0 & 0 & 0 & b & 0 \\
 0 & 0 & -b & 0 & 0 & 0 & 0 & b
\end{array} \right)
\end{equation}
has eigenvalues $0$, $b$, $2b$, and 
$\left( 1+2b \pm \sqrt{(1+2b)^2-2b(3+b)} \right)\Big/2$
so that it is always non-negative. Note that the spectrum
of $\lambda_{AB}$ is the same as the spectrum of the partial
transpose $\rho_{AB}^{T_A}$ (cf.~~\cite{bib_horo_one}), as expected.
This confirms that the condition based on 
$\Lambda=\Gamma\otimes I$ and Peres' separability condition 
are equivalent for $2 \times n$ systems (when $\Gamma$ is applied
to the two-dimensional system and $I$ to the $n$-dimensional one).
In this example, applying the dual map ${\tilde \Lambda} = I\otimes \Gamma$
yields a positive operator which traces to 3.

\end{document}